\documentclass[%
 reprint,
 nofootinbib,
 nobibnotes,
 amsmath,amssymb,
 aps,mathrsfs,
 twocolumn,
 pre,
]{revtex4-1}

\usepackage{multirow}         
\usepackage{tabularx}         %
\usepackage[roman]{parnotes}  %
\makeatletter
\def\parnoteclear{%
    \gdef\PN@text{}%
    \parnotereset
}
\makeatother

\usepackage{graphicx}
\usepackage{subcaption}

\usepackage{epstopdf}
\usepackage{color}
\graphicspath{{./img/}}

\DeclareGraphicsExtensions{.ps}
\usepackage{float}
\usepackage{booktabs}
\usepackage{dcolumn}
\usepackage{bbm}
\usepackage{bm}
\usepackage{chngcntr}

\newcommand{\bfr}{\mathbf{r}}
\newcommand{\cs}{c_{\rm s}}

\newcommand{\eq}{\mbox{\tiny{eq}}}

\newcommand{\sigl}{\sigma_{\rm l}}
\newcommand{\sigs}{\sigma_{\rm s}}
\newcommand{\rhol}{\rho_{\rm l}}
\newcommand{\rhos}{\rho_{\rm s}}

\newcolumntype{Y}{>{\centering\arraybackslash}X}
\DeclareMathOperator{\Tr}{Tr}

\begin{document}


\title{Phase diagrams and crystal--fluid surface tensions in additive and nonadditive two--dimensional hard disk mixtures}

\author{Shang-Chun Lin}
 \email{shang-chun-lin@uni-tuebingen.de}
\affiliation{%
 Institut 
 f\"ur Angewandte Physik, Universit\"at T{\"u}bingen, Auf der Morgenstelle 10, 72076 T{\"u}bingen,Germany
}%
\author{Martin Oettel}
 \email{martin.oettel@uni-tuebingen.de}
\affiliation{%
 Institut 
 f\"ur Angewandte Physik, Universit\"at T{\"u}bingen, Auf der Morgenstelle 10, 72076 T{\"u}bingen,Germany
}%

\date{\today}

\begin{abstract}
Using  density functionals from fundamental measure theory, phase diagrams and 
crystal--fluid surface tensions in additive and 
nonadditive (Asakura-Oosawa model) two-dimensional hard disk mixtures are determined
for the whole range of size ratios $q$ between disks, assuming random disorder in the crystal phase. 
The  fluid--crystal transitions are first--order due to the assumption of a periodic unit cell in the density functional
calculations. Qualitatively, the shape of the phase diagrams is similar to the case of three--dimensional hard sphere mixtures. 
For the nonadditive case, a broadening of the fluid--crystal coexistence region is found for small $q$ whereas for higher $q$
a vapor--fluid transition intervenes. In the additive case, we find a sequence of spindle type, azeotropic and eutectic phase diagrams upon lowering $q$ from 1 to 0.6. The transition from azeotropic to eutectic is different from the three--dimensional
case. Surface tensions in general become smaller (up to a factor 2) upon addition of a second species and they are rather small.
The minimization of the functionals proceeds without restrictions and optimized graphics card routines are used.

\end{abstract}

\pacs{123456789}
\maketitle


\section{\label{sec:Intro}Introduction}

The fluid--crystal transition in two--dimensional (2D) systems of hard disks has been of fundamental interest over the past years.
Only recently, it has been established in the one--component system by simulations \cite{bernard2011two} and experiments
\cite{thorneywork2017two}  that the transition happens
via a first--order transition from the fluid to the hexatic phase and a continuous transition from the hexatic to the crystal 
phase. Although the crystal phase is not strictly periodic (it does not have infinitely long--ranged positional order),
in simulations and experiments it has practically the appearance of a conventional, periodic crystal. 
Therefore, 2D hard disks have a similar status as a model system for crystallization in films and monolayers as 3D hard spheres 
have for crystallization in the bulk. Besides simulations, classical density functional theory (DFT) for hard particle systems
has reached a certain maturity and accuracy owing to the development of fundamental measure theory,
starting with the work of Rosenfeld \cite{rosenfeld1989free}. For 2D hard disks, a functional has been proposed
in Ref.~\cite{roth_2d} which gives a very accurate description of fluid structure in one-- and two--component systems
\cite{Thorneywork2014}, as well as values for the fluid and crystal coexistence densities which are rather close
to the ones of the first--order fluid--hexatic transition \cite{roth_2d}. In these FMT calculations, strict periodicity
of the crystal phase was assumed. 
 
Crystals in binary hard disk systems have been studied some time ago by ``older'' density functional methods
in Refs.~\cite{Oxtoby1990,Baus1990} (variants of weighted--density functionals with restricted minimizations).
For substitutionally disordered crystals, a sequence of phase diagram types (spindle, azeotropic, eutectic)
has been found upon lowering the disk size ratio similar to the case of 3D hard spheres \cite{Frenkel1991}, although
the exact shape and the transition size ratios differ considerably between Refs.~\cite{Oxtoby1990,Baus1990}.
In these cases, the crystals were assumed strictly periodic as well. In Ref.~\cite{Likos1993}, a survey of possible alloy 
phases was undertaken in a special zero--temperature limit (identifying the highest packing fraction structure among all candidates).
 
In last decade, phase field crystal (PFC) models have emerged as an efficient tool to phenomenologically describe  
phase diagrams of binary systems in 2D and 3D \cite{elder2007phase, greenwood2011modeling} and of the crystal--fluid 
surface tension \cite{provatas2010phase}. The PFC models employ a certain Taylor expansion of direct correlation functions among 
species to produce the desired crystal structure and use several parameters to capture material properties. 
Whereas the approach is suited to 
to describe the mesoscale behavior of solidification generically, a link to the density distributions in specific crystals
is difficult to establish. An example is the hard sphere system where PFC fails to describe quantitatively vacancy concentrations,
surface tensions and associated density profiles \cite{oettel2012description}.       

Only recently, a binary mixture with a fixed size ratio of $1/1.4$ was investigated by simulations aiming at the fate of the
hexatic phase \cite{russo2017disappearance} in disordered crystals. The hexatic phase was found to disappear quickly upon
addition of the smaller species; overall, a phase diagram of eutectic type was found for this size ratio.     

Here, we employ the FMT functional of Ref.~\cite{roth_2d} to study phase diagrams and crystal--fluid surface tensions
for additive and nonadditive binary hard disk mixtures. The nonadditive case is the 2D variant \cite{AO2d} of the well--known Asakura--Oosawa (AO) model
\cite{asakura1954interaction,asakura1958interaction}, originally formulated for a mixture of 3D hard spheres where there
are no interactions between particles of the second component (depletant). The depletants lead to an effective attractive
potential between particles of the first species (which for small size rations is strictly a two--body potential),
therefore the study of the AO model is equivalent to a study of hard disks with additional short--ranged attractions.
The derivation of the AO functional from the functional of Ref.~\cite{roth_2d} proceeds by employing the
``linearization trick'' already studied in the 3D case \cite{schmidt2000density,mortazavifar2016fundamental}.  
We examine the case of random disorder over the whole range of 
possible size ratios. Random disorder includes the cases of substitutional disorder when disk sizes are comparable, interstitial disorder
for small size ratios and superpositions of alloy configurations for intermediate size ratios.

The paper is organized as follows. In section \ref{sec:theory}, we introduce the theoretical background  for the AO model, 
FMT and the FMT-based AO functional. In section \ref{sec:numric}, we discuss the numerical treatment of the 
full minimization of the functionals for bulk crystals and crystal--fluid surfaces. In section \ref{sec:result}, 
we present our results for density distributions in the crystal and crystal--fluid interfaces, for phase diagrams and surface tensions. 
In the final section, we summarize and discuss our results.

\section{\label{sec:theory}Theory}  

\subsection{\label{subsec:AO} Hard disk mixture and Asakura--Oosawa model}

We consider a mixture of large (l) and small (s) disks, with diameter $\sigl$ and $\sigs$, respectively, and
$q=\frac{\sigs}{\sigl}$ denoting the size ratio.
In the case of an additive system (denoted as HD mixture), one may define an interaction diameter $d_{ij}=\sigma_{i}/2 + \sigma_j/2$ with
$i,j=\{{\rm l}, {\rm s}\}$. The pair potential $\Phi^{ij}(r)$ between two particles with center-center distance $r$ is
$\infty$ for $r<d_{ij}$ and 0 for $r>d_{ij}$. 

In the case of an AO mixture, the interaction diameter $d_{\rm ss}$ for the interaction between two small disks is zero, i.e.
there is no interaction among the small disks and they behave as an ideal gas. The other interaction diameters $d_{\rm sl}$
and $d_{\rm ll}$ remain unchanged.
The small disks act as an depletant and induce an effective, attractive two--body potential $\Phi^{\tiny{\mbox{AO}}}(r)$ 
between the large disks
(depletion potential). Its shape is determined by the overlap of exclusion areas of two large disks at distance $r$,
where the exclusion areas (which are forbidden to centers of small disks) are circles of diameter $\sigl+\sigs$ centered at the midpoints of the 
large disks \cite{asakura1954interaction,AO2d}:
\begin{equation}
\beta\Phi^{\tiny\mbox{AO}}(r)=\begin{cases}
	\infty\hspace{5pt}\mbox{if}\,\hspace{5pt}r<\sigl\\
	-\eta^{\prime}\left[\cos^{-1}\left(\frac{r}{\sigl\left(1+q\right)}\right)-\frac{r}{\sigl\left(1+q\right)}\sqrt{1-(\frac{r}{\sigl\left(1+q\right)})^2}\right] \\
	\hspace{100pt}\mbox{if}\,\hspace{5pt}\sigl<r<\sigl+\sigs\\
	0\hspace{5pt}\mbox{otherwise}     ,          
\end{cases}
\label{eqn:AO_potential}
\end{equation}
where $\eta^{\prime}=\frac{\sigs^2}{2}\rhos\left( \frac{1+q}{q} \right)^2$ determines the magnitude of the depletion
potential ($\rhos$ is the bulk number density of small disks).  Furthermore, 
$\beta=\frac{1}{k_{\tiny\mbox{B}}T}$, with $k_{\tiny\mbox{B}}$ denoting Boltzmann's constant, and  $T$ temperature. 
For small size ratios $q < \frac{\sigs}{\sigl} \leq \frac{2-\sqrt{3}}{\sqrt{3}}\simeq 0.155$, the AO mixture can be mapped 
exactly onto a single component model with an effective two--body potential given by the depletion potential above. For larger $q$, 
the effective potential should include $n$--body overlaps of excluded area ($n \geq 3$). 
Furthermore, in the dilute limit of the (additive) HD mixture (with the number density $\rhol$ of large disks being small),
the effective potential between large disks is identical to Eq.\eqref{eqn:AO_potential} \cite{AO2d}.
 
\subsection{\label{subsec:FMT}Fundamental Measure theory (FMT)}
We consider inhomogeneous mixtures with density profiles $\rho(\bfr)=\left\lbrace\rhos(\bfr),\rhol(\bfr)\right\rbrace$. 
In classical density functional theory (DFT), crystals are considered as inhomogeneous fluid, whose equilibrium density profile 
$\rho_{\eq}(\bfr)=\left\lbrace\rho_{\rm s,\eq}(\bfr),\rho_{\rm l,\eq}(\bfr)\right\rbrace$ minimizes the grand potential 
\begin{equation}
\label{eqn:grand_potential}
\Omega[{\rho}]=F[\rho]-\sum_{i}\int\,d\mathbf{r}\,\rho_{i}(\bfr)(\mu_i-V_i^{\mbox{\tiny{ex}}}(\bfr)),
\end{equation}
where $F$ is the Helmholtz free energy, $\mu_i$ and $V_i^{\mbox{\tiny{ex}}}$ are the chemical potential and 
the external potential for species $i$, respectively. $F$ can be further decomposed into the ideal gas part 
$F_{\tiny{\mbox{id}}}$ and the excess free energy $F_{\tiny{\mbox{ex}}}$. The exact form of $F_{\tiny{\mbox{id}}}$ is 
\begin{equation}
\label{eqn:free_idea_gas}
\beta F_{\mbox{\tiny{id}}}=\sum_{i=\{{\rm s,l}\}}\int\,d\mathbf{r}\,\rho_i\left(\mathbf{r}\right)\left[ \ln\left(\rho_i\left(\mathbf{r}\right)\lambda_i^2\right)-1\right] \hspace{1em}
\end{equation}
where $\lambda_i$ is the thermal wavelength for species $i$. In the following, we put $\lambda_i=1$.

FMT is the most accurate route to density functionals of hard body mixtures. Most FMT functionals
assume an excess free energy density which is local in a set of weighted densities $n_{\alpha}(\bfr)$
which are convolutions of the density profiles with geometrically motivated weight functions 
\cite{rosenfeld1989free}. For hard spheres in 3D, the original derivation of the functionals proceeds from
an exact low--density form (``deconvolution of the Mayer $f$--bond'') and subsequently uses scaled particle arguments
\cite{rosenfeld1989free}. Such a functional does not describe crystals, though. In this case, a possible derivation
proceeds via dimensional crossover. Here, one requires that by confining an arbitrary density profile to
1D (a line) and 0D (a collection of points), the functional delivers the correct free energies whose
exact form are known from other arguments \cite{Tarazona1997,Tarazona2000}. Using this route, the properties of hard--sphere crystals
and crystal--fluid interfaces are described in quantitative agreement with simulations 
\cite{oettel2010free,Haertel2012,oettel2012description}. Also, the low--density form remains exact.

In the derivation of a genuine 2D functional along these lines, problems are encountered. Maintaining the
exact low--density form or the having the exact free energy for a density distribution consisting of two sharp peaks
is not possible with an excess free energy density local in weighted densities \cite{Tarazona1997}. An approximate solution
to this problem was derived in Ref.~\cite{roth_2d}. The excess free energy is given by
\begin{equation}
 \beta F_{\mbox{\tiny{ex}}}^{\mbox{\tiny{HD}}}[n_\alpha]= \int \, d\bfr\, \Phi^{\tiny{\mbox{HD}}}({n_{\alpha}})
\end{equation}
where the weighted densities $n_\alpha$ are sums over convolutions of the HD species density profiles with weight functions,
\begin{eqnarray}
\label{eqn:weighted_density}
n_{\alpha}(\mathbf{r})&=&\sum_{i=\{{\rm s,l}\}} \int\,d\mathbf{r}^{\prime}\,\rho_i(\mathbf{r}^{\prime})w_{\alpha}^i(\mathbf{r-r}^{\prime}) \\
   &=:& n_{\alpha}^{\rm s}+n_{\alpha}^{\rm l} \nonumber
\end{eqnarray} 
where $\alpha$ indicates the type of weight function and $i$ the species (l=large and s=small). 
The weight functions are defined as 
\begin{eqnarray}
\label{eqn:weight_functions}
&w_1^i(\bfr)=\delta(R_i-r),\hspace{5pt} &w_0^i(\bfr)=\frac{\delta(R_i-r)}{2\pi R_i},\nonumber\\
&w_2^i(\bfr)=\Theta(R_i-r),\hspace{5pt} &\mathbf{w}_1^i(\bfr)=\frac{\bfr}{r}\delta(R_i-r),\nonumber\\
&\left(\mathbf{w}_{\mbox{\tiny T}}^i(\bfr)\right)_{\alpha\beta}=(\frac{r_\alpha r_\beta}{\bfr^2})\delta(R_i-r),&\nonumber\\
\end{eqnarray}
where $R_i$ is the radius of species $i$, $\theta(r)$ is the Heaviside step function, and $\delta(r)$ is the Dirac delta function. 
$\mathbf{w}_{\mbox{\tiny T}}^i(\bfr)$ is a tensorial weight function with cartesian components $\alpha\beta$.
 \\
The free energy density is given by \cite{roth_2d}: 
\begin{align}
\label{eqn:free_energy_functional_HS}
&\Phi^{\tiny{\mbox{HD}}}(n_{\alpha})=\nonumber\\
&-n_0\ln(1-n_2)+\frac{\left(C_0 n_1^2 +C_1 {\bf{n}}_1^2 + C_2 \Tr[{\bf{n}}_T^2] \right)}{4\pi(1-n_2)} \;,
\end{align}
with three parameters $C_0$, $C_1$ and $C_2$.
The functional gives the correct second virial coefficient if $C_0 + C_2/2 = 1$. Furthermore, the correct free energy for a 
single, sharp density peak requires $C_0 + C_1 + C_2 = 0$. Thus, the dependence on the three parameters can be reduced to a
dependence on a single parameter $a$ with
\begin{equation}
C_0 = \frac{a+2}{3}, \hspace{5pt} C_1 = \frac{a-4}{3} \hspace{5pt}\mbox{and}\hspace{5pt} C_2 = \frac{2-2a}{3}. 
\end{equation}
For one component, a best fit to the Mayer $f$--bond gives $a=11/4$ whereas a fit to crystal pressures obtained by simulation
gives $a=3$ \cite{roth_2d}. For binary systems in the fluid phase, the functional delivers an excellent description of pair correlation
functions when compared to experiments \cite{Thorneywork2014}. 

Recently, a functional for 2D rods (discorectangles) has been derived which maintains the exact low--density form by using
weighted densities which are two--center convolutions with a weight function (fundamental mixed measure theory, FMMT)
\cite{Wittmann2017}. In the limit of the 2D rods becoming disks, the functional (\ref{eqn:free_energy_functional_HS})
is an approximation to the FMMT functional. However, fluid--crystal coexistence densities in the one--component case
are approximately equal, and the numerical effort in FMMT is considerably higher. Therefore we will not consider
the FMMT functional in this work.

A functional for the AO mixture can be obtained by the ``linearization recipe'': A functional for a genuine hard--body mixture
(such as the one in Eq.~(\ref{eqn:free_energy_functional_HS})) is linearized in the density (or equivalently in the weighted densities
$n_\alpha^{\rm s}$) of the small species. This entails that the direct correlation function between two particles of the 
small species, $c^{(2)}_{\rm ss}(\mathbf{r},\mathbf{r}')=-\beta\delta^2 F_{\tiny\mbox{ex}}/(\delta\rhos(\mathbf{r})\delta\rhos(\mathbf{r}'))$, 
vanishes, consistent with the small species behaving as an ideal gas. In 3D, such a functional (derived
from the original Rosenfeld functional \cite{rosenfeld1989free}) describes structural properties
and wetting transitions in the fluid phase very well \cite{Brader2003}. Recently, an extension using functionals
from the dimensional crossover route has been studied which allows the description of the 
crystal phase in 3D \cite{mortazavifar2016fundamental}.  According to the linearization recipe, the AO mixture excess free
energy density is given by
\begin{eqnarray}
\label{eqn:f_AO_ex}
\Phi^{\tiny{\mbox{AO}}}\left(\left\lbrace n_{\alpha}^{\rm l},n_{\alpha}^{\rm s} \right\rbrace\right)=&\Phi^{\tiny{\mbox{HD}}}\left( n_{\alpha}^{\rm l}\right)+\sum\limits_{\alpha}n_{\alpha}^{\rm s}\frac{\partial\Phi^{\tiny{\mbox{HD}}}\left( n_{\alpha}^{\rm l} \right)}{\partial n_{\alpha}^{\rm l}} \;.
\end{eqnarray} 

\section{\label{sec:numric}Numerical methods}
\subsection{\label{subsec:numeric_free}Free minimization and phase coexistence }

\begin{figure}
\includegraphics[width=0.5\textwidth]{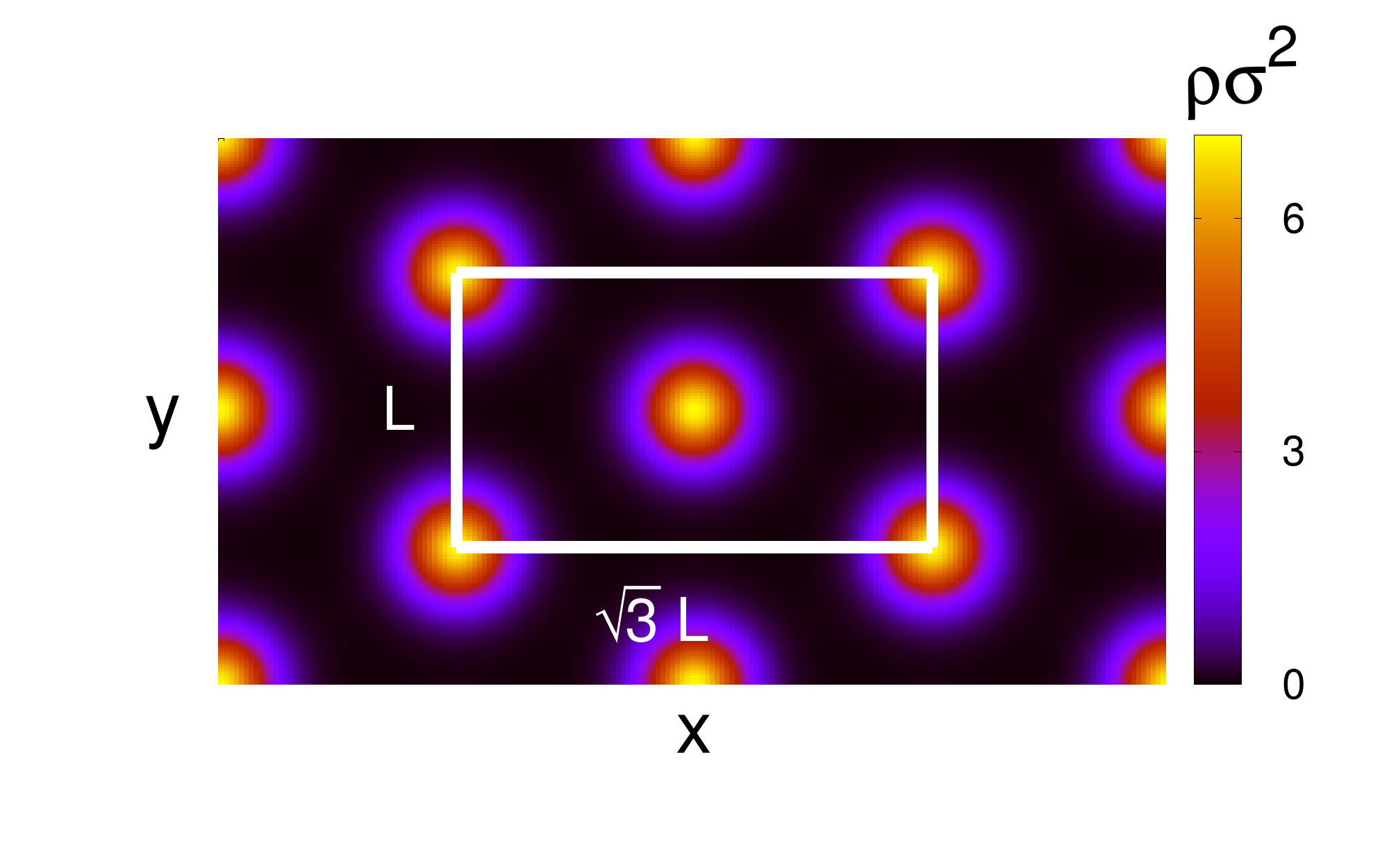}
        \caption{Density distribution $\rho(\bfr)$ of a one--component perfect crystal ($\bar{\rho}\sigma^2=0.93$ and $a = 11/4$).
 The solid white line indicates the computational box (rectangular unit cell of the triangular lattice which contains 
two particles)}
\label{fig:unit_cell}
\end{figure}

For the crystal phase, we assume periodicity and consider a rectangular unit cell 
with side lengths $L$ and $\sqrt{3}L$ for a triangular lattice 
(see Fig.\ref{fig:unit_cell}). Since we only consider solid solutions (random disorder), we 
assume that the triangular lattice is formed by equilateral triangles as in the one--component case. 

The free parameters in this free energy minimization problem are the density profiles
$\rhol(\mathbf{r})$ and $\rhos(\mathbf{r})$ in the unit cell as well as the length $L$. We parametrize
the latter via an effective vacancy concentration $n$:
\begin{equation}
\int_{\mbox{\tiny{cell}}} d\bfr\, (\rhol(\bfr)+\rhos(\bfr)) =: 2(1-n)=(\bar{\rho}_{\rm l}+\bar{\rho}_{\rm s}) \sqrt{3}\,L^2
\end{equation}
In the one--component case, an ideal crystal has 2 particles in the unit cell, therefore $n>0$ indeed
corresponds to the vacancy concentration in the equilibrium crystal. For a HD mixture, $n$ may also be negative,
corresponding to an effective interstitial concentration which is easily possible if small disks are inserted
into a crystal of large disks.
  
The full minimization for given average densities $\bar{\rho}_{\rm l}, \bar{\rho}_{\rm s}$ proceeds via
\begin{equation}
F_{\mbox{\tiny{eq}}}(\bar{\rho}_{\rm l}, \bar{\rho}_{\rm s})
  =\min\limits_{n} \min\limits_{\left\lbrace\rhol(\bfr),\rhos(\bfr)\right\rbrace}F[n_\alpha],
\end{equation} 
i.e. in two steps \cite{oettel2010free}. The first minimization step is
achieved by an iterative solution of the Euler--Lagrange equation (for fixed $n,L$)
\begin{equation}
\rho_i=\exp\left(-\beta\frac{\delta F_{\tiny{\mbox{ex}}}\left[n_\alpha\right]}{\delta\rho_i}+\beta\mu_i\right)=K[\rho_i],
\end{equation} 
where
\begin{equation}
\label{eqn:dFdn}
\beta\frac{\delta F_{\tiny{\mbox{ex}}}\left[{n_\alpha}\right]}{\delta\rho_i(\bfr)}=\int\,d\mathbf{r^{\prime}}\sum_{\alpha}\frac{\partial\Phi [n_{\alpha}]}{\partial n_{\alpha}^i}(\bfr') w_{\alpha}^{i}(\bfr^{\prime}-\bfr)
\end{equation}
where $\Phi$ is given by Eq.\eqref{eqn:free_energy_functional_HS} in case of the HD mixture, 
and Eq.\eqref{eqn:f_AO_ex} in case of the AO mixture. 
The chemical potentials $\mu_i$ are adapted in each iteration step to keep $\bar{\rho}_{\rm l}, \bar{\rho}_{\rm s}$ constant. 
Iteration is done using a combination of Picard steps and DIIS (discrete inversion in iterative subspace) \cite{oettel2012description,oettel2010free}. The Picard steps are performed according to
\begin{equation}
\rho_i^{j+1}=\xi \,K[\rho_i^j]+(1-\xi)\rho_i^j,
\end{equation}
where $i$ is the species index, $j$ labels the iteration step and $\xi$ is a Picard mixing parameter which we chose
in the range from $10^{-3}$ to $10^{-2}$ for bulk crystal and also interface minimizations. 
The DIIS steps are performed using between 5 and 9 forward profiles. As a side remark, we also minimized $F[n_\alpha]$ 
by dynamic DFT with the exponential time differencing algorithm \cite{cox2002exponential} for a one--component system. 
By choosing the time step $dt = 10^{-3}$ (in units of the Brownian time), the thermodynamic properties in equilibrium are identical 
to the ones from the Picard--DIIS method, but the dynamic DFT method requires much more computational resources.\\
The second minimization step, the minimization with respect to $n$ (and thus $L$), amounts to do the first minimization for
a few values of $n$ within an interval of starting width $\sim 10^{-3}$
and determine the minimum via a quadratic fit. The procedure is iterated with smaller interval widths until we have reached 3 digits of confidence or the interval width is less than $10^{-5}$.


The procedure is slightly modified in the case of an AO mixture, see also Ref.\cite{mortazavifar2016fundamental} for more details. 
Here, we define the vacancy concentration by
\begin{equation}
\int_{\mbox{\tiny{cell}}} d\bfr\,\rhol(\bfr)  =:2(1-n)=\bar{\rho}_{\rm l} \sqrt{3}\,L^2 \;,
\end{equation}
i.e. it corresponds to the concentration of sites unoccupied by the large particles. Furthermore, we define
a semi--grand free energy (fixed $\bar{\rho}_{\rm l}$ and $\mu_{\rm s}$)
\begin{equation}
  F^{\prime}=F-\mu_{\rm s} \int\,d\bfr\,\rhos(\bfr)
\end{equation}
which is minimized in step 1 for fixed $n,L$. In each iteration step, the density profile $\rhos(\bfr)$ of the small spheres
is computed by the grand--canonical equilibrium condition which can be solved explicitly: 
\begin{eqnarray}
\label{eqn:f_AO_ex_linear}
&\frac{\delta\Omega[\rhol,\rhos]}{\delta \rhos} \stackrel{!}= 0 \Rightarrow \nonumber\\
&\rhos(\mathbf{r}) = \exp\left(\beta\mu_{\rm s}-\int\,d\mathbf{r}^{\prime}\sum\limits_\alpha \frac{\partial\Phi_{\tiny{\mbox{ex}}}^{\tiny{\mbox{HD}}}\left[ n_{\alpha}^l \right]}{\partial n_{\alpha}^{\rm l}}w_{\alpha}^{\rm s}(\bfr^{\prime}-\bfr)\right) \;. \nonumber\\
\end{eqnarray}

Phase coexistence requires $P_{\rm cr}=P_{\rm fl}$ and $\mu_{\rm i,cr}=\mu_{\rm i,fl}$ with $i = l,s$, i.e.
coexisting fluid [crystal] states form two lines in the $\rhol$--$\rhos$ plane. 
In practice, first we choose $\rhol$ and $\rhos$ for the crystal and treat $\rhol=\rho_{\rm l,cr}$ as the parameter
on which the other three coexistence densities depend. 
Fully minimizing $F/N$ with $n$ delivers $P_{\rm cr}$ and $\mu_{\rm i,cr}$.
Through  $\mu_{\rm i,cr}=\mu_{\rm i,fl}$ and the fluid equation of state we can find $P_{\rm fl}$, $\rhol$, $\rhos$ 
in the fluid. In general, $P_{\rm fl} \neq P_{\rm cr}$ and thus we change $\rho_{\rm s,cr}$ iteratively until
$|P_{\rm cr}-P_{\rm fl}|<5\times10^{-6}$.
\subsection{\label{subsec:gamma}Surface tension}

A surface tension in 2D is a line tension defined as $\frac{\Omega+P A}{L}$, where $P$ is the pressure, $A$ is the area of the system and 
$L$ is the length of the interface. In this paper, we are interested in the planar surface tension $\gamma$, 
which is determined by the slope of the free energy density versus the inverse length of the numerical box in the direction
of the interface normal,  with the average particle density fixed \cite{adland2013phase}.  

In general, $\gamma$ depends on the angle $\theta$ between the crystal and the interfacial normal. 
For small anisotropies, $\gamma$ can be approximated by $\gamma(\theta)=\gamma_0(1+\epsilon \sin(6\theta))$; 
in FMT, $\epsilon \simeq O[10^{-3}]$ for the one--component crystal-fluid interface. 
In experiments \cite{thorneywork2017two}, $\epsilon\simeq O[10^{-2}]$. Due to the smallness of $\epsilon$, in this paper, $\gamma_0$ is directly determined by $\frac{\gamma(0)+\gamma(\pi/6)}{2}$.

The density profiles are initialized similar to Ref.~\cite{oettel2012description}.
In the iterations we chose a Picard mixing parameter constant in space (this works here in 2D but not 
in 3D ~\cite{oettel2012description}).  We fix the average densities 
$\bar{\rho_i}=\frac{\bar{\rho}_{i,\mbox{\tiny{cr}}}+\bar{\rho}_{i,\mbox{\tiny{fl}}}}{2}$ by adapting $\mu_i$ in the iterations, 
where $\bar{\rho}_{i,\mbox{\tiny{cr/fl}}}$ is the bulk average density in the crystal/fluid phase 
for species $i$ at coexistence, and then finally perform the free minimization.   

\subsection{\label{subsec:numeric_detail}Further numerical details}
Here we briefly discuss further computational details. The crystal phase requires double precision, with numbers of grid points 
from $64^2$ up to $256^2$ for one unit cell. The crystal--fluid interfaces require an extension of the numerical box between 
1 $\times$ 96 and 1 $\times$ 196 unit cells to give reliable surface tensions. 
Heavy usage of Fourier transforms is required for the minimization.
Weighted densities (Eq.~\eqref{eqn:weight_functions}) are computed using
\begin{equation}
\mathfrak{F}(n_\alpha^i)=\mathfrak{F}(\rho^i)\mathfrak{F}(w_\alpha^i)
\end{equation} 
and functional derivatives (Eq.~\eqref{eqn:dFdn}) by
\begin{equation}
\mathfrak{F}\left(\frac{\delta F_{\tiny{\mbox{ex}}}\left[{n_\alpha}\right]}{\delta\rho_i}\right)=\sum_{\alpha}\mathfrak{F}\left(\frac{\partial\Phi_{\tiny{\mbox{ex}}}[n_{\alpha}]}{\partial n^{i}_{\alpha}}\right)\mathfrak{F^{\star}}(w_\alpha^i)
\end{equation}
with $\mathfrak{F}$ denoting the Fourier transform and $^{\star}$ the complex conjugate. For $\mathfrak{F}(w_\alpha)$, 
the analytic forms using Bessel functions are used \citep{GPU_FMT}. 
For accelerating the numerics, all calculations are executed on high-performance Nvidia Tesla K80 or K40 GPU's 
with massive parallelization through the developer environment CUDA \cite{CUDA}. 
For a detailed description of GPU utilization in two-- and three--dimensional FMT,  we refer the reader to the paper by Stopper 
\emph{et al.} \cite{GPU_FMT}. CUDA has a wide range of tools and libraries, such as template library thrust and fast Fourier transforms (cuFFT) 
which is usually a bottleneck in the DFT calculations. With a potential speed gain of up to 40 times relative to a serial CPU program
\cite{GPU_FMT}, our calculations gave a factor of 15--20 since we try to maximize the system size; thus, our largest system is 4 times 
larger than those in Ref.\cite{GPU_FMT}. The minimization of a unit cell (first minimization step) usually takes a few seconds 
($\sim 500$ Picard-DIIS steps) and that of an interface about 15--30 minutes ($\sim 5000$ Picard-DIIS steps) for one--component system. 

\section{Results}
\label{sec:result}

\begin{table}[t]
\centering
\caption{Thermodynamic properties of the one--component crystal--fluid transition. 
$\gamma_0$ denotes the averaged planar surface tension, $\sigma$ the HD diameter, $\mu$ chemical potential, $P$ pressure, 
$\eta=(\pi/4)\sigma^2 \rho$ packing fraction, superscript (${\rm 1c}$) one--component, 
and subscripts (${\rm co}$) coexistence of the crystal  (${\rm cr}$) and fluid  (${\rm fl}$), respectively. 
Note that for Exp and MC, two values for $\eta_{\rm cr}$ correspond to 
the packing fraction of the hexatic phase at fluid-hexatic coexistence and the packing fraction at
the hexatic--crystal continuous transition, respectively. The FMT coexistence values for $a=11/4$
differ slightly from those in Ref.~\cite{roth_2d} which suffer from a small numerical error. }
\label{tab:co_HD}
\begin{tabularx}{0.5\textwidth}{@{}YYYYY@{}}
 \hline \hline \\
\parnoteclear
              			& \multicolumn{2}{c}{FMT} & \multirow{2}{*}{Exp}\cite{thorneywork2017two}$^*$ & \multirow{2}{*}{MC}\cite{bernard2011two}\\
 						& $a=11/4$       & $a=3$      &                     &                      \\\hline \\
$\beta \sigma\gamma_0^{\rm 1c}$ 				& 0.0992 	&	0.0815    	&0.1 		&   \\[2mm]
$\beta \sigma^2 P_{\rm co}^{\rm 1c}$      			& 10.84 	&  	9.234   &  			&   9.185\\[2mm]
$\beta\mu_{\rm co}$    			& 14.576	&  	12.778  &    		&	\\[2mm]
$\eta_{\rm cr}^{\rm 1c}$   			& 0.732		&  	0.7165  &0.7/0.73		&	0.716/0.72  \\[2mm]
$\eta_{\rm fl}^{\rm 1c}$   			& 0.711		&  	0.6913  &0.68		&	0.700\\[2mm] 
\hline\hline
\end{tabularx}
($^*$ see Supplementary Material in Ref.~\cite{thorneywork2017two})
\parnotes
\end{table}

\subsection{One--component system}

In the last decade, two--dimensional one--component HD systems were extensively studied, with now quantitative
agreement in type and location of the phase transition between experiments \cite{thorneywork2017two} and Monte Carlo (MC) simulations 
\cite{bernard2011two}. For a summary, in Table~\ref{tab:co_HD} we provide a comparison between FMT, MC and experiments. 
Experiments and simulations find a first--order transition between the fluid and the hexatic phase, and a continuous
transition between the hexatic and crystal phase.
The surface tension in the experiments \cite{thorneywork2017two} (see Supplementary Material therein) is for hexatic---fluid coexistence. FMT results are for an assumed first--order transition between fluid and crystal.

From Table \ref{tab:co_HD} we see that coexistence packing fractions and the surface tension are described very well by FMT, even 
though in FMT the strict periodicity assumption for  the crystal differs from the character of the hexatic and crystal phase in
experiments/simulations. This good correspondence is in line with the quantitative description of fluid structure given in 
earlier work \cite{roth_2d,Thorneywork2014}.

\begin{figure*}[t]
    \begin{subfigure}[bl]{0.4\textwidth}
        \includegraphics[width=\textwidth]{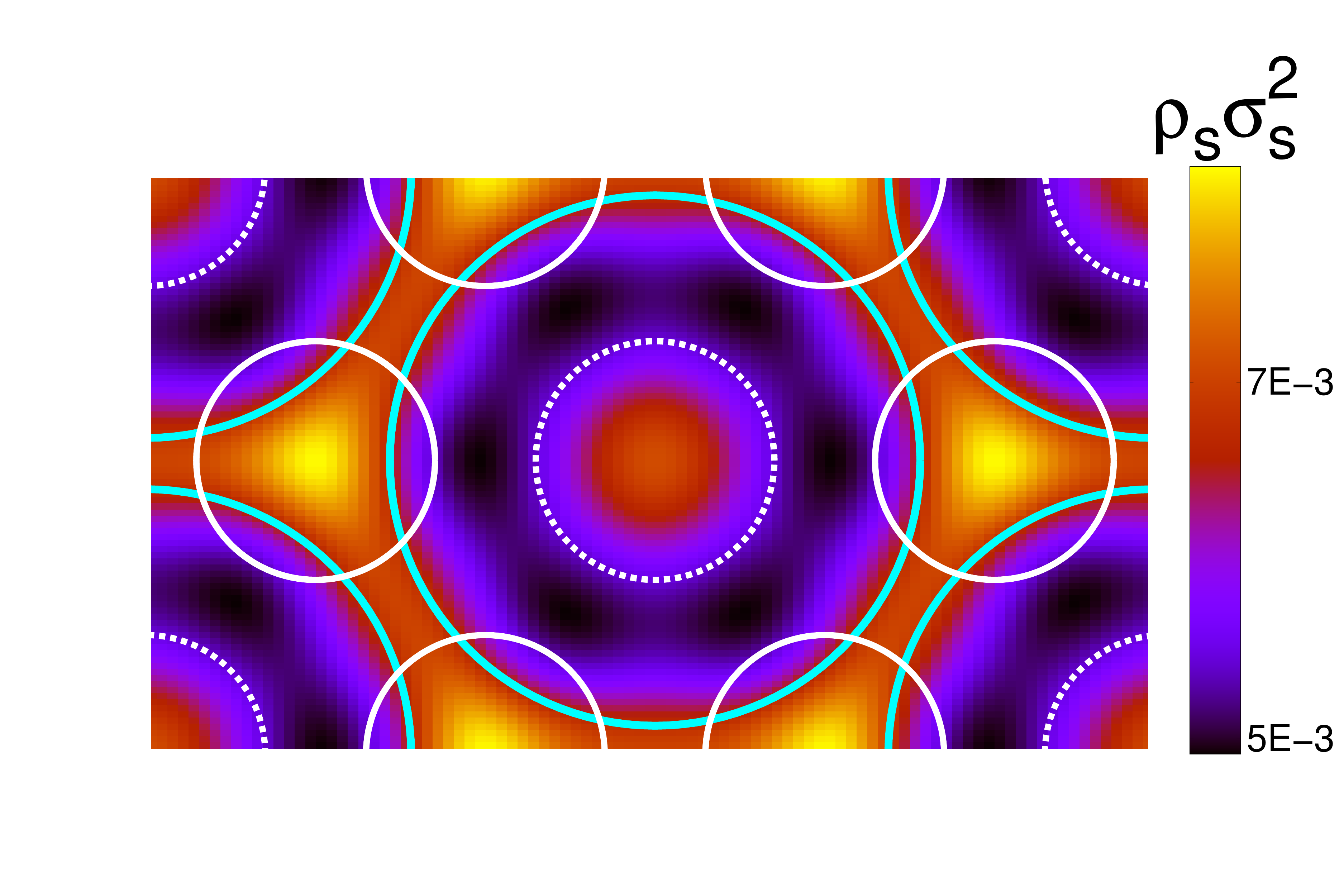}
        \subcaption{$\cs=0.03$}
        \label{fig:example_q_0.45_s_low_unit}
    \end{subfigure}
    \begin{subfigure}[bl]{0.4\textwidth}
        \includegraphics[width=\textwidth]{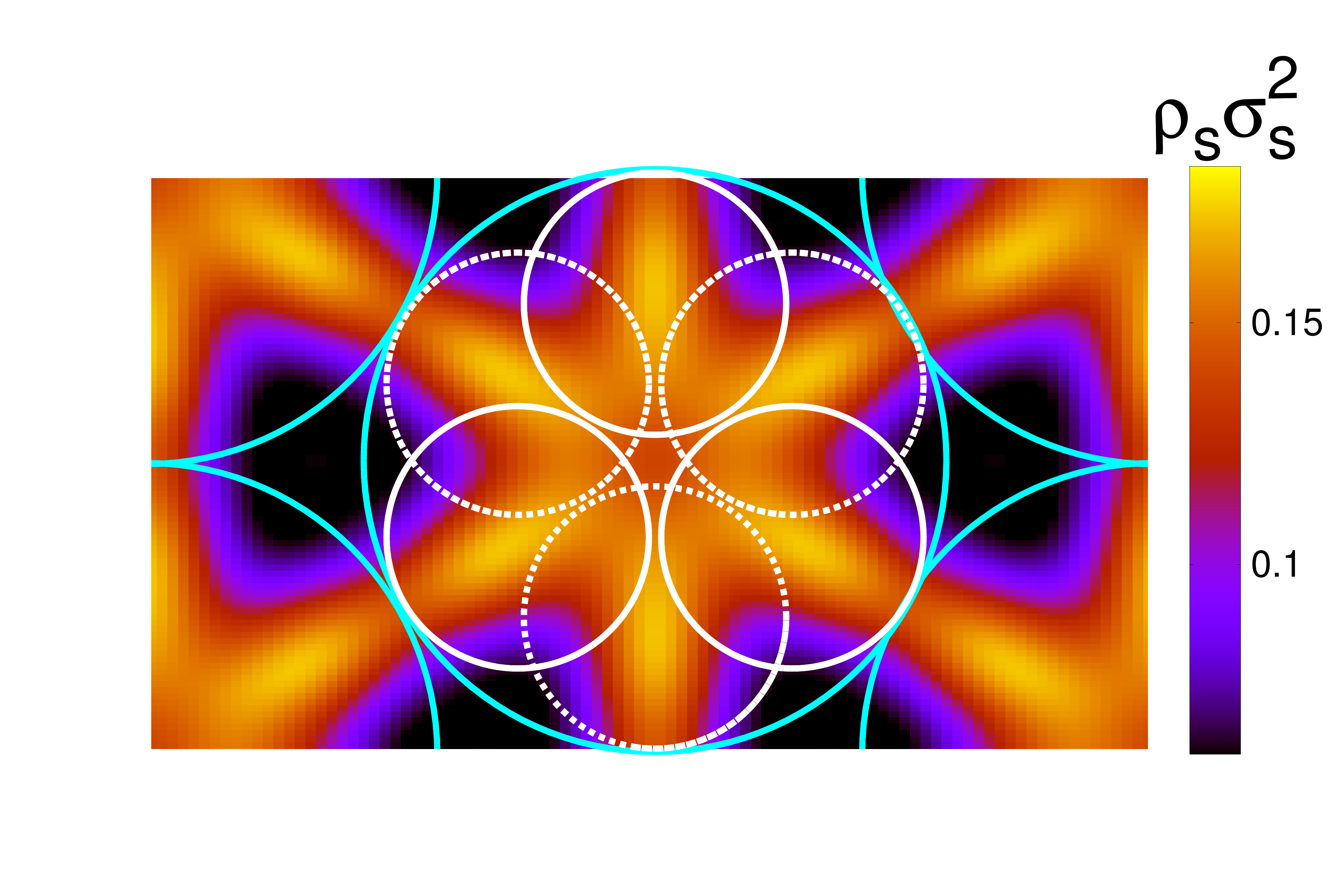}
        \subcaption{$\cs=0.39$}
        \label{fig:example_q_0.45_s_unit}
    \end{subfigure}
    \caption{Density profiles for small disks in a HD mixture crystal with $q=0.45$ at crystal--fluid coexistence for small disk concentration $\cs=0.03$ (a)
and $\cs=0.39$ (b). The solid blue circles indicate the extension of large disks, solid white and dashed circles indicate the extension of small disks.  In both cases, 
the density profile is a superposition of substitutional and interstitial disorder. In (b), interstitial disorder dominates and is compatible with
an AB$_3$ alloy structure where one large disk is replaced by three small disks (white solid or dashed circles). }
    \label{fig:example_unit_cell}
\end{figure*} 

\subsection{Binary systems: crystal density profiles}
\label{sec:crystaldensity}

When the radii of the disks are comparable (large $q \lesssim 1$), we observe a clear substitutional disorder. Density peaks for both species are centered
on the triangular lattice points and their magnitude is essentially determined by the composition of the crystal. An example can be seen in the crystal part
of the crystal--fluid density profile shown in Fig.~\ref{fig:example_q_0.75_s} below. 

For small size ratios $q \ll 1$, we observe interstitial disorder, i.e. the small disks almost exclusively occupy the interstitial space between the large disks
which in turn occupy the triangular lattice points.
An example can be seen in the crystal part
of the crystal--fluid density profile shown in Fig.~\ref{fig:example_q_0.15} below. The HD and AO case are very similar, and qualitatively the AO crystal density
profiles in 3D show the same behavior \cite{mortazavifar2016fundamental}. 

For intermediate $q$ and the HD case, we observe a superposition of substitutional and interstitial disorder, and the interstitial disorder may show a transition to different alloy configurations upon changing the composition. We exemplify this for $q=0.45$. Large disks density peaks are again centered
on the triangular lattice positions (not shown).  For low small disk concentrations ($\cs=0.03$) we observe interstitial disorder superficially compatible with an AB$_2$ structure  (see the small disk distribution in Fig.~\ref{fig:example_q_0.45_s_low_unit}). From the large and small disks drawn in Fig.~\ref{fig:example_q_0.45_s_low_unit} one sees however that the small disks are too big for the formation of a true AB$_2$ phase. For higher small disk concentrations ($\cs=0.39$, see Fig.~\ref{fig:example_q_0.45_s_unit} for the small disk distribution) the lattice constant becomes smaller (large spheres on the triangular lattice points almost touch) and the interstitial density peaks of the small spheres   are compatible with an AB$_3$ structure. Here, remarkably, the large disks drawn around the triangular lattice points and the small disks drawn around the interstitial peak positions reveal two packed AB$_3$ configurations.
In the AO case, we only observed small disk density distribution of the type shown in  Fiq.~\ref{fig:example_q_0.45_s_low_unit}. 

Here, we have not investigated whether the minimized crystal structures with disorder are stable or not with respect to phase separation into different alloy 
phases. This requires more extensive investigations beyond the scope of this work. However, our results illustrate that a free minimization of the 
FMT functional is capable of generating alloy structures without the need to explicitly parameterize the density profiles (e.g. by suitably chosen Gauss peaks,
as it is often done).  

\begin{figure*}
    \begin{subfigure}[b]{0.4\textwidth}
        \includegraphics[width=\textwidth]{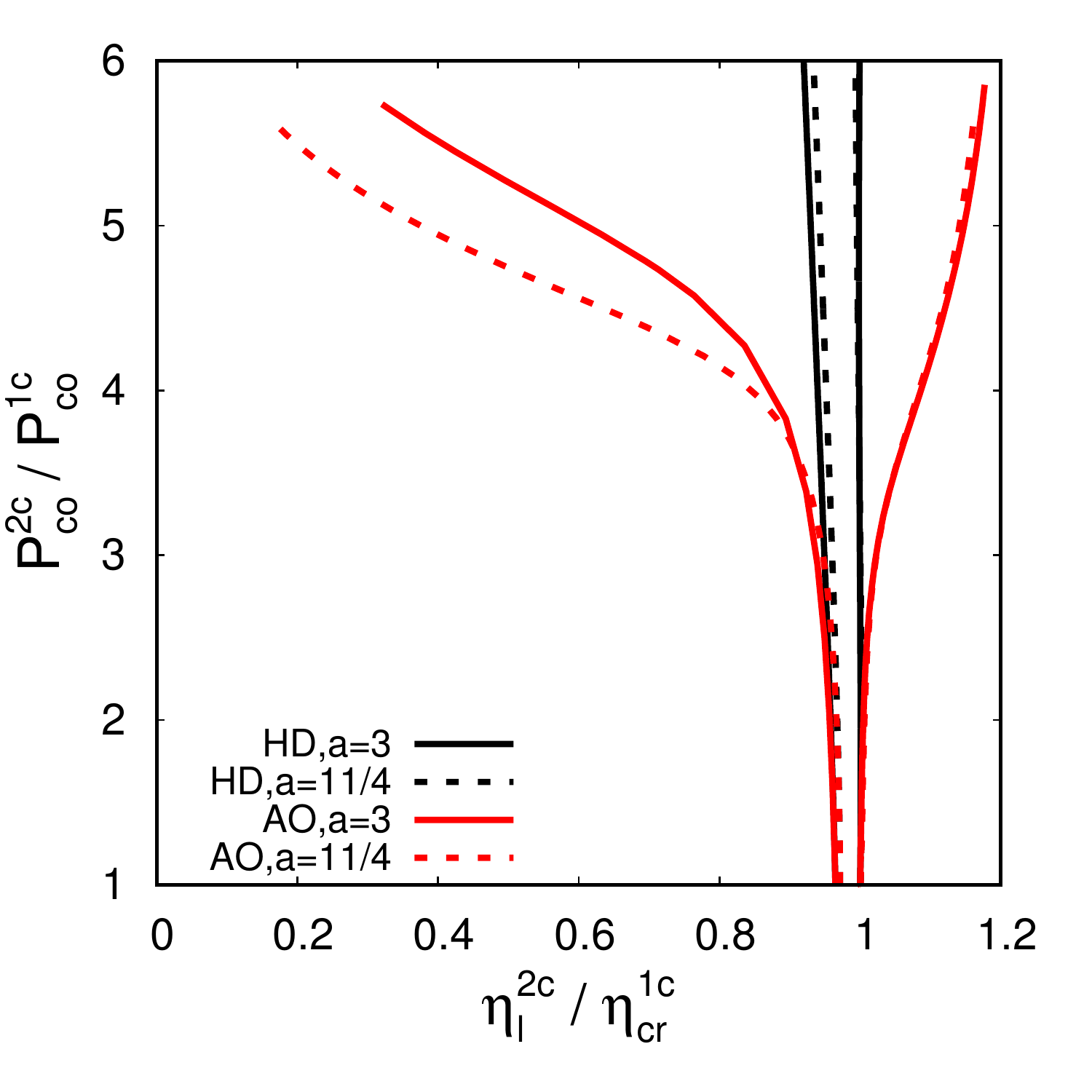}
        \caption{}
        \label{fig:phase_q_0.15_etal}
    \end{subfigure}
    \begin{subfigure}[b]{0.4\textwidth}
        \includegraphics[width=\textwidth]{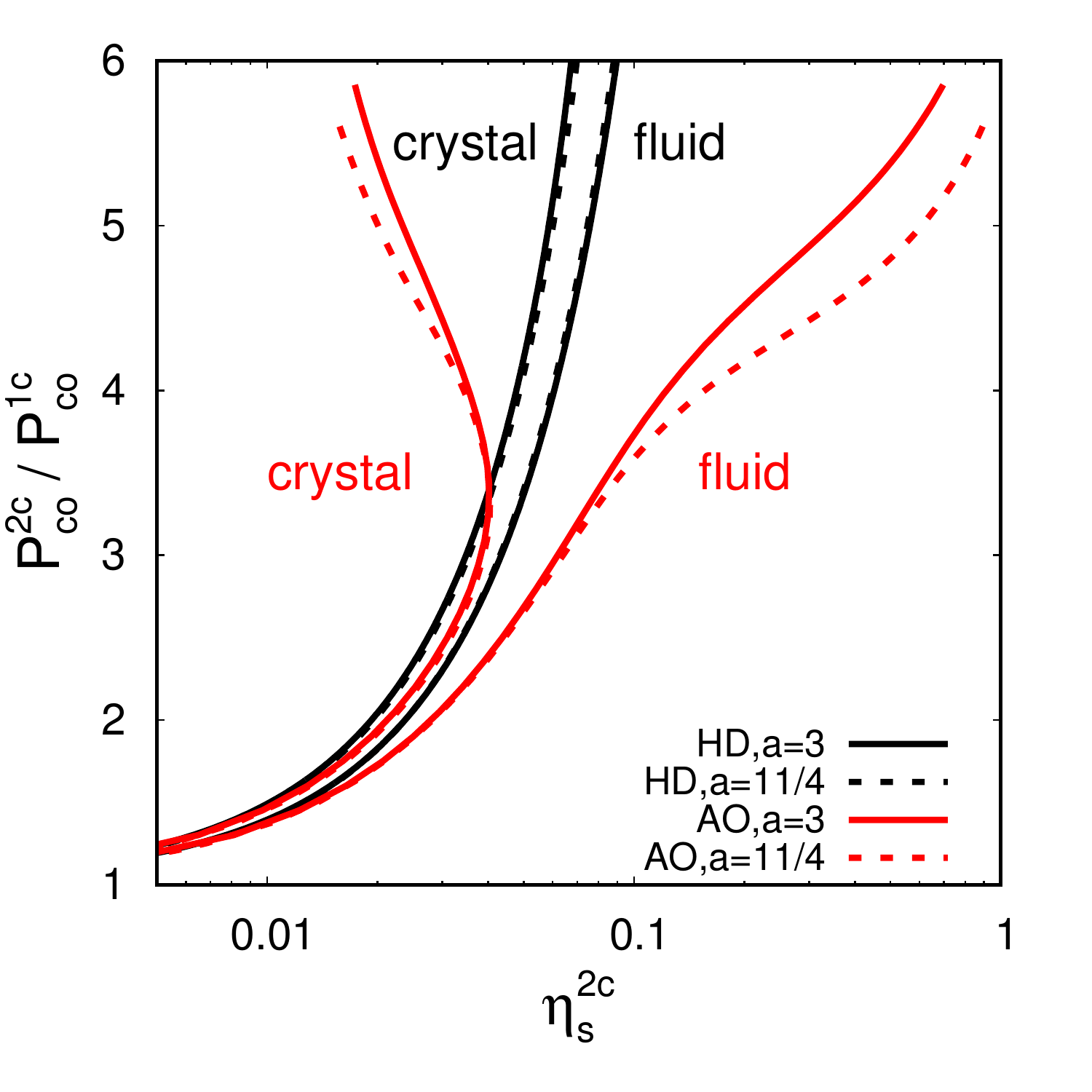}
        \caption{}
        \label{fig:phase_q_0.15_cs}
    \end{subfigure}
\caption{Phase diagram for the size ratio $q=0.15$ in (a) the $\eta_{\rm l}$--$P$ plane and in (b) the
$\eta_{\rm s}$--$P$ plane. In (a), pressure and packing fraction are normalized by the coexistence values
of the 1--component HD system $P_{\rm co}^{\rm 1c}$ and $\eta_{\rm cr}^{\rm 1c}$. 
 }
\label{fig:phase_q_0.15}
\end{figure*}

\begin{figure*}[t]
     \begin{subfigure}[b]{0.4\textwidth}
        \includegraphics[width=\textwidth]{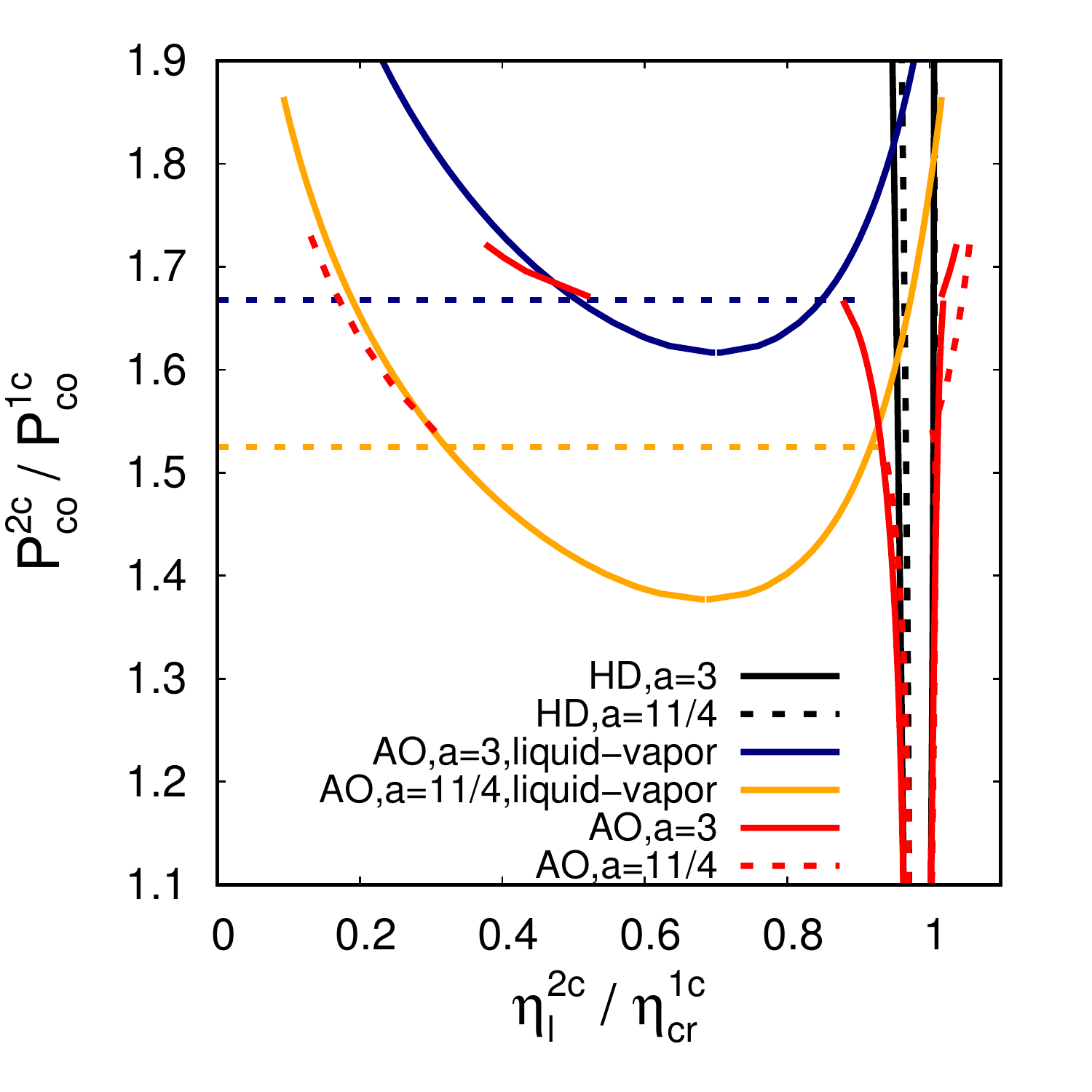}
        \caption{$q=0.3$}
        \label{fig:phase_q_0.3}
    \end{subfigure}
     \begin{subfigure}[b]{0.4\textwidth}
        \includegraphics[width=\textwidth]{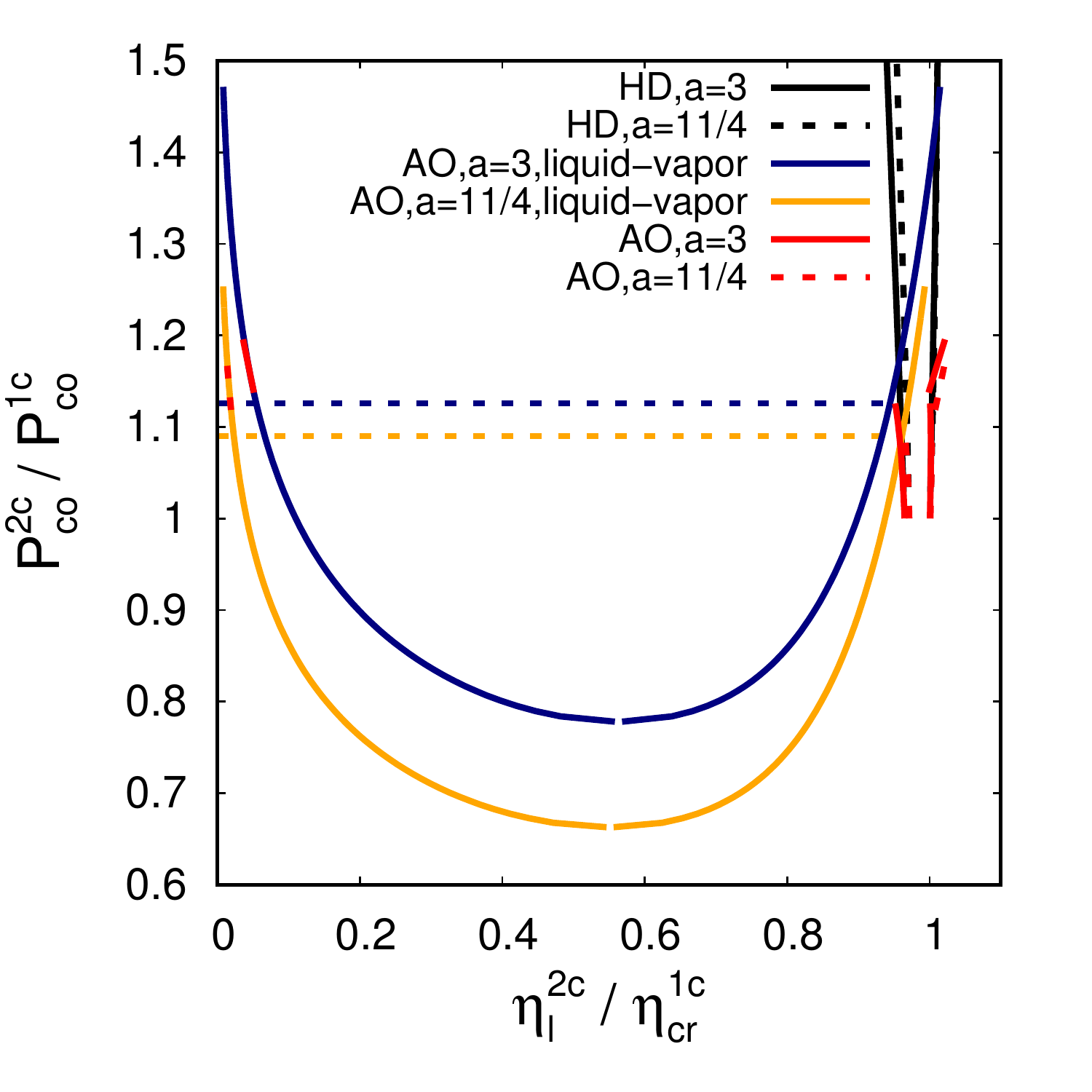}
        \caption{$q=0.45$}
        \label{fig:phase_q_0.45}
    \end{subfigure}
    \caption{Binary mixture phase diagrams for $q=0.3$ (a) and $q=0.45$ (b)
in the $\eta_{\rm l}$--$P$ plane. Pressure and packing fraction are normalized by the coexistence values
of the 1--component HD system $P_{\rm co}^{\rm 1c}$ and $\eta_{\rm cr}^{\rm 1c}$. The triple point pressures
are indicated by horizontal dotted lines.
}
    \label{fig:phase_diagrams_smallq}
\end{figure*}

\begin{figure*}[t]
    \begin{subfigure}[b]{0.32\textwidth}
        \includegraphics[width=\textwidth]{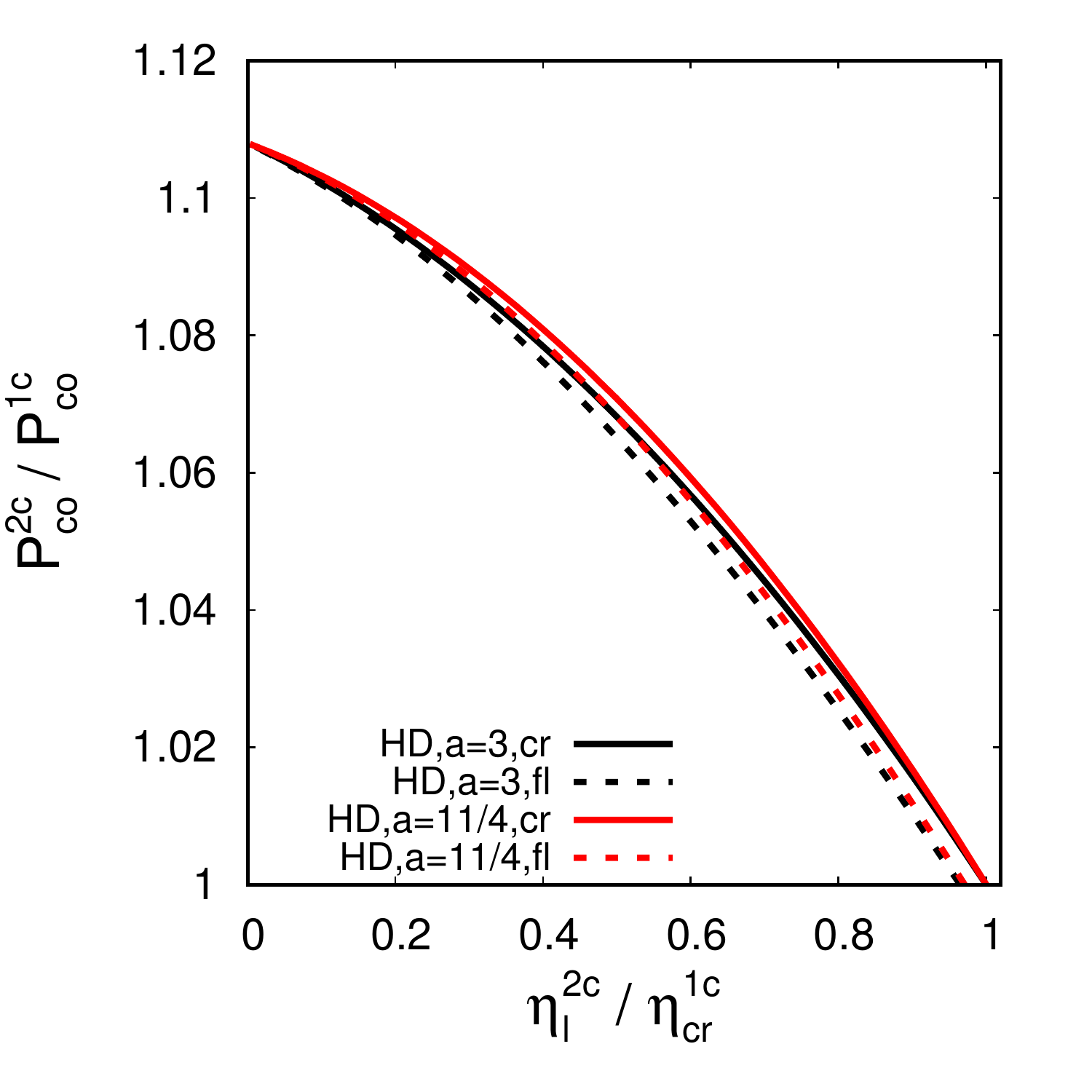}
        \caption{$q=0.95$}
        \label{fig:phase_q_0.95}
    \end{subfigure}
     \begin{subfigure}[b]{0.32\textwidth}
        \includegraphics[width=\textwidth]{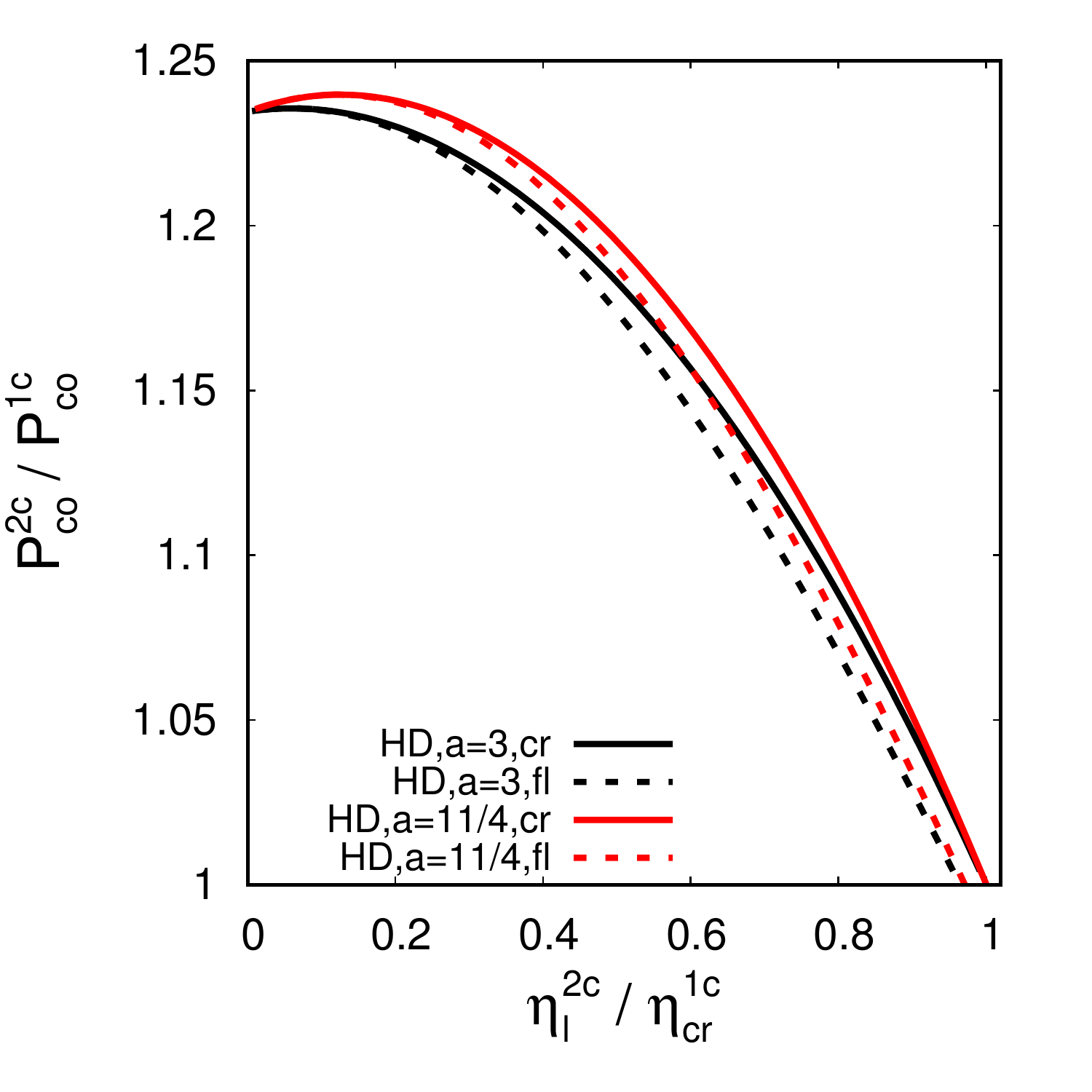}
        \caption{$q=0.9$}
        \label{fig:phase_q_0.9}
    \end{subfigure}
     \begin{subfigure}[b]{0.32\textwidth}
        \includegraphics[width=\textwidth]{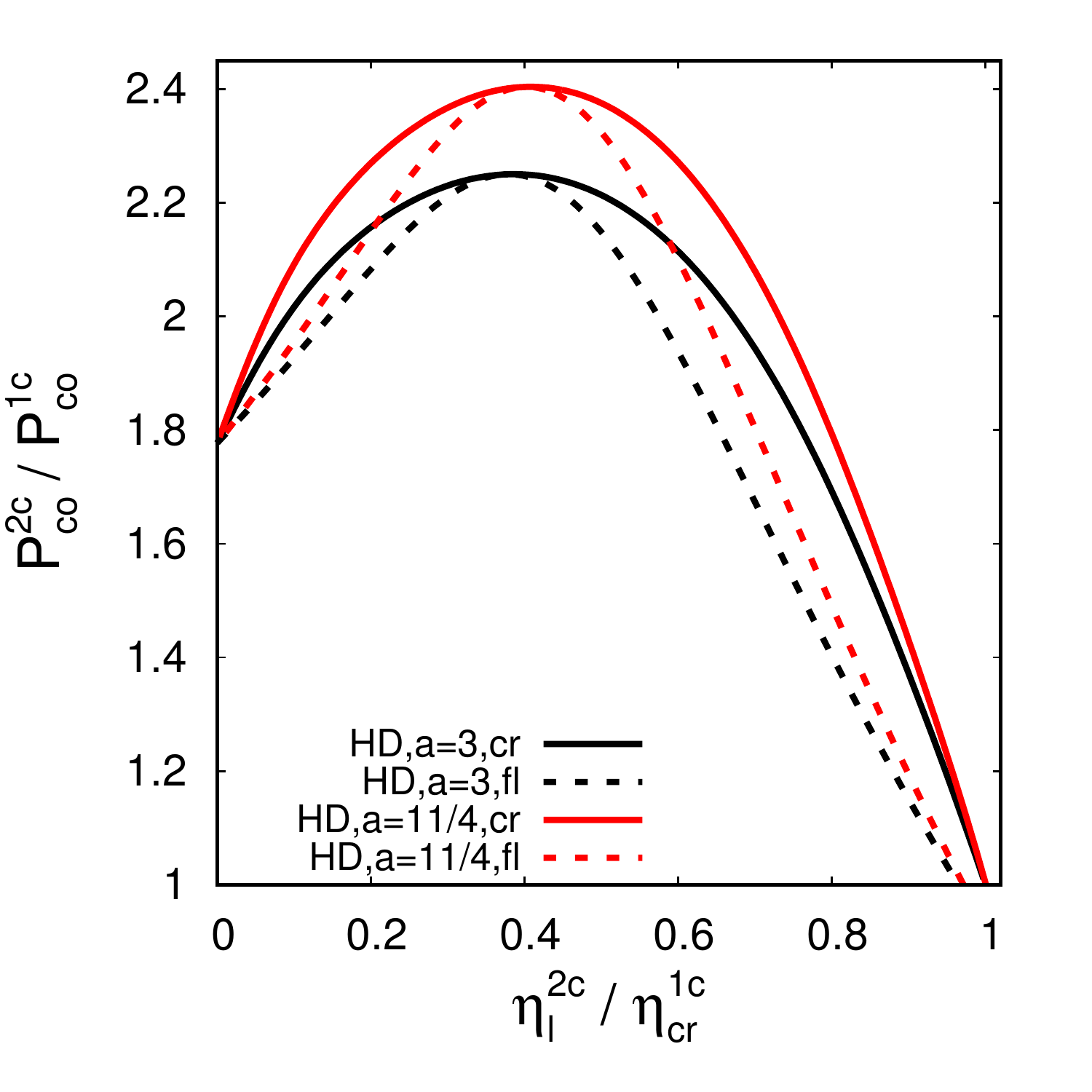}
        \caption{$q=0.75$}
        \label{fig:phase_q_0.75}
    \end{subfigure}
    \caption{Binary mixture phase diagrams for $q=0.95$ (a), $q=0.9$ (b) and $q=0.75$ (c) 
in the $\eta_{\rm l}$--$P$ plane. Pressure and packing fraction are normalized by the coexistence values
of the 1--component HD system $P_{\rm co}^{\rm 1c}$ and $\eta_{\rm cr}^{\rm 1c}$. 
}
    \label{fig:phase_diagram_big_q}
\end{figure*}

\begin{figure*}[t]
    \begin{subfigure}{0.32\textwidth}
        \includegraphics[width=\textwidth]{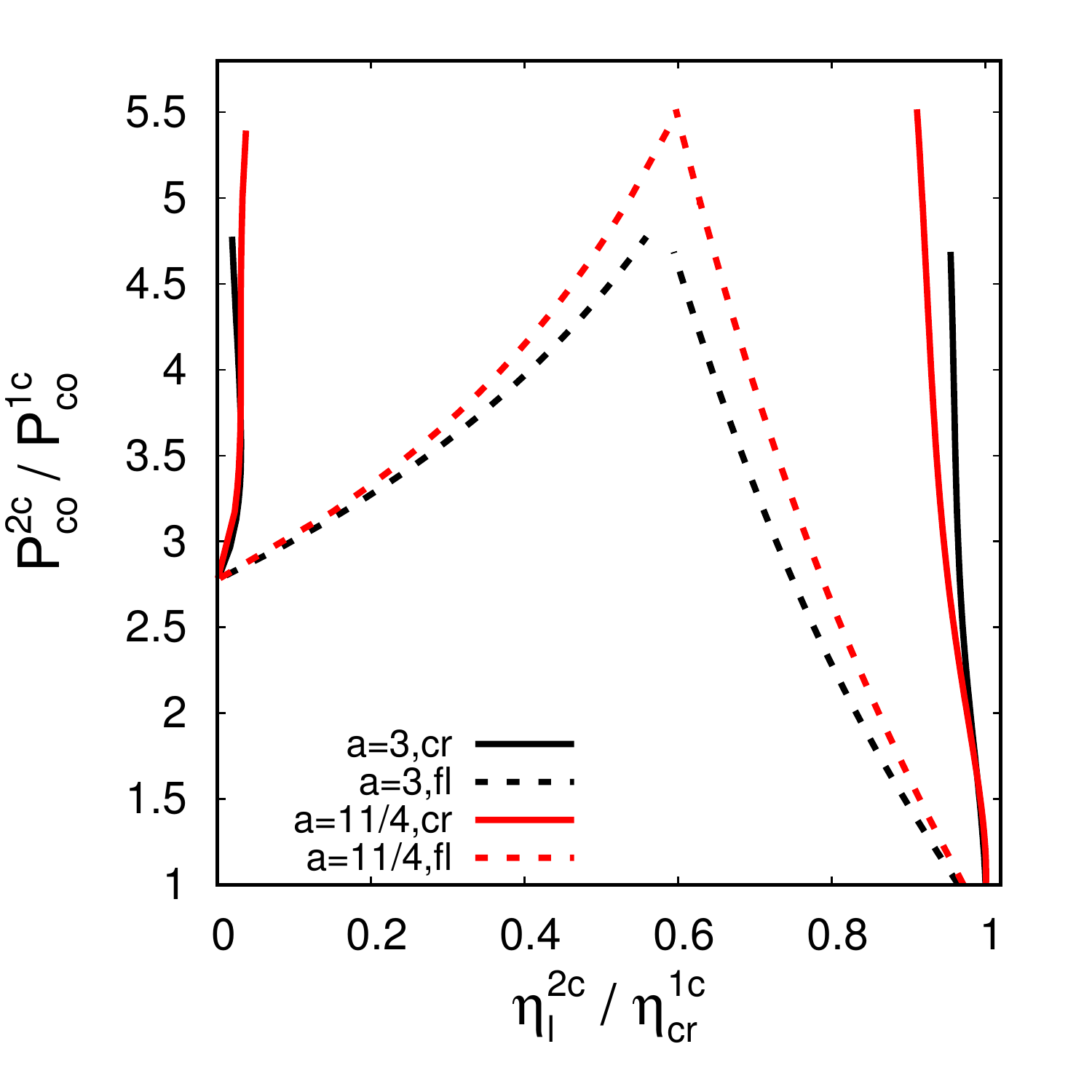}
        \caption{$q=0.6$}
        \label{fig:phase_q_0.6_eu}
    \end{subfigure}    
    \begin{subfigure}{0.32\textwidth}
        \includegraphics[width=\textwidth]{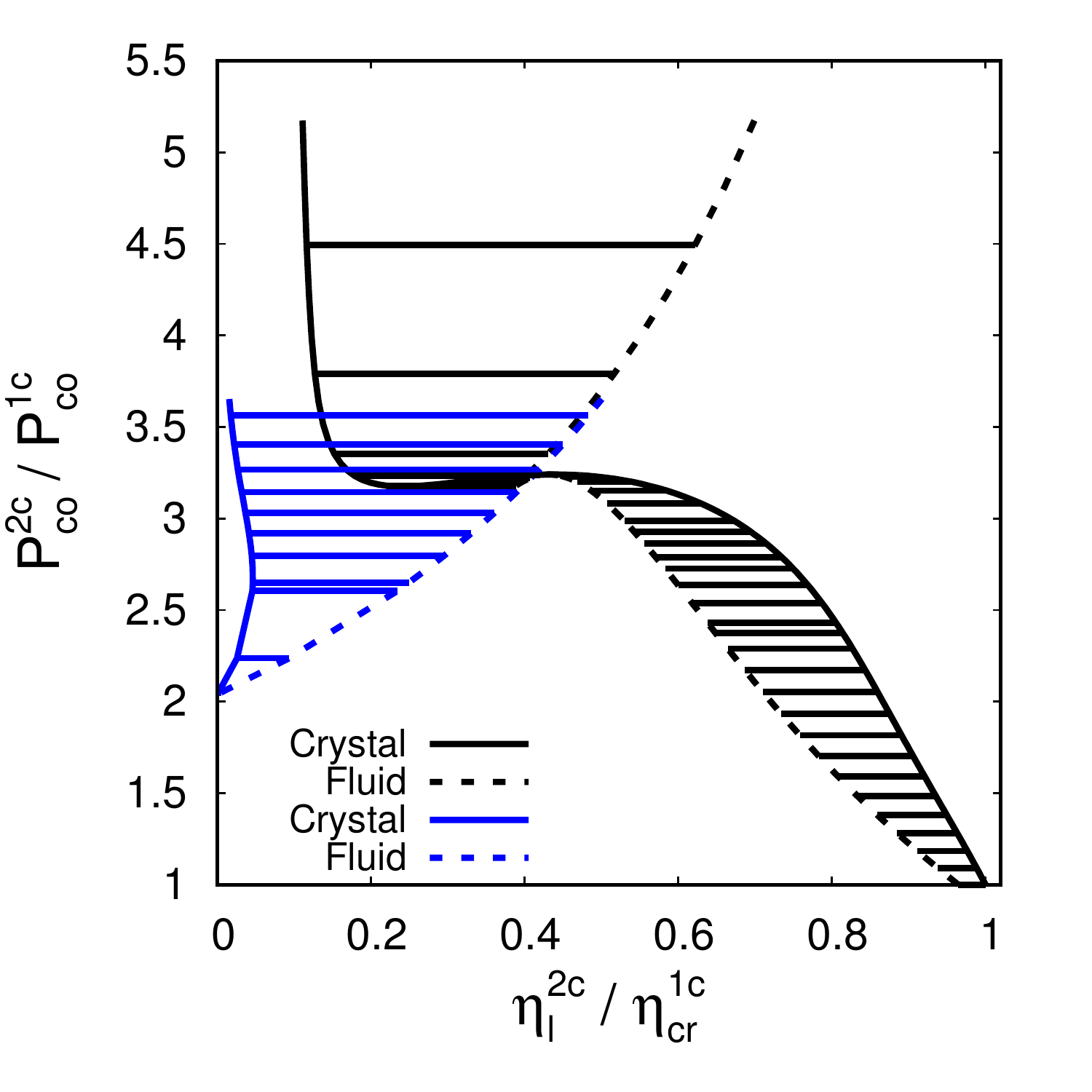}
        \caption{$q=0.7$}
        \label{fig:phase_q_0.7_eu}
    \end{subfigure}  
        \begin{subfigure}{0.32\textwidth}
        \includegraphics[width=\textwidth]{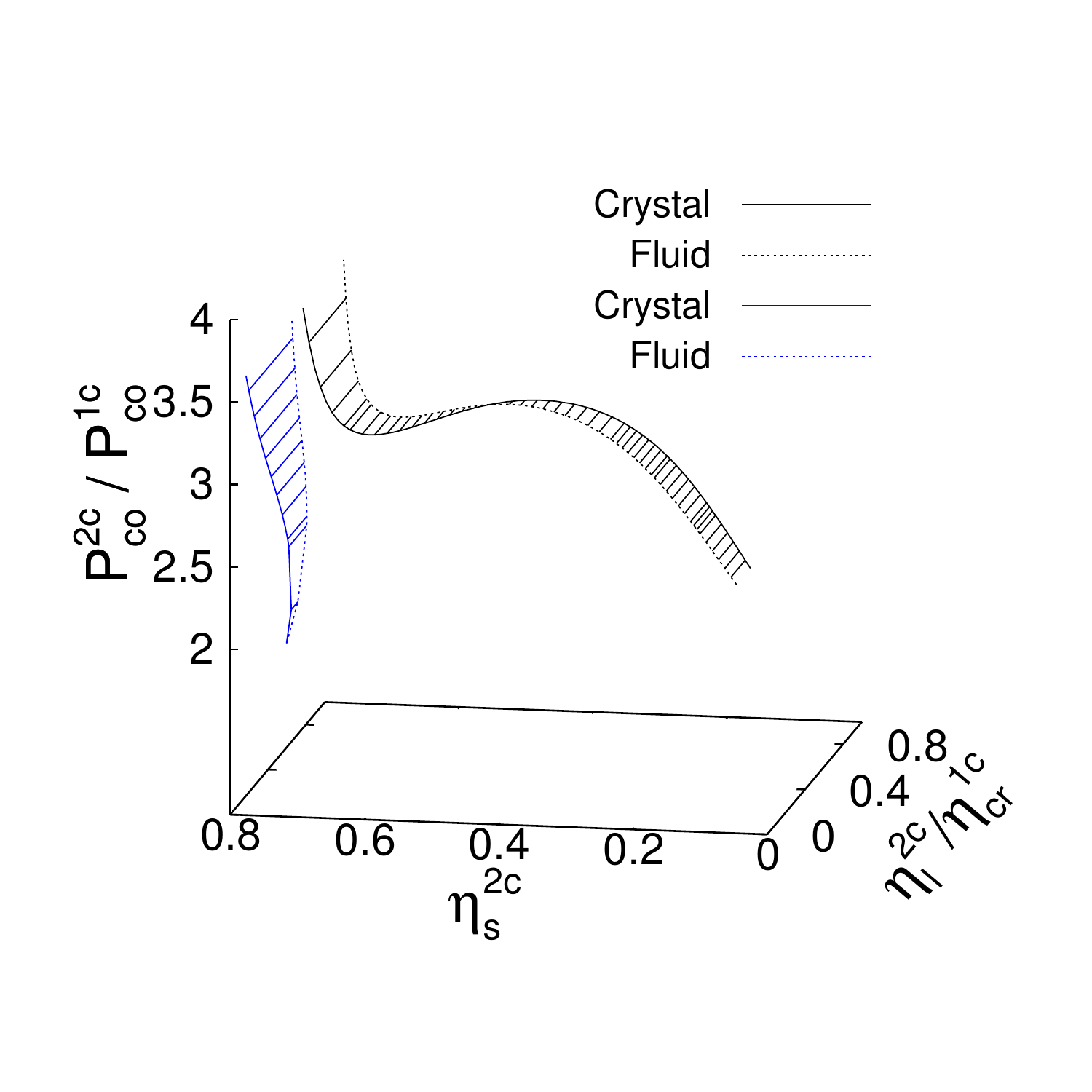}
        \caption{$q=0.7$ in $\eta_{\rm l}$--$\eta_{\rm s}$--$P$ space}
        \label{fig:phase_q_0.7_eu_3d}
    \end{subfigure}   
    \caption{Binary mixture phase diagrams for $q=0.6$ (a) and $q=0.7$ (b), (c). Only the results for $a=3$ are shown in (b), (c). }
    \label{fig:phase_diagram_intermediate_q}
\end{figure*}

\subsection{Binary systems: phase diagrams}

For two--component hard systems, equilibrium states are on a surface in a three--dimensional space, spanned by e.g.
the packing fractions
$\eta_{\rm l}$, $\eta_{\rm s}$ and the pressure $P$. Consequently, binodals are lines in this three--dimensional space and they are
often displayed by their two--dimensional projections, e.g. lines in the $\eta_{\rm l}$--$P$ or $c_{\rm s/l}$--$P$ plane
where $c_{\rm s/l}=\rho_{\rm s/l}/(\rhol+\rhos)$ is the relative concentration of small/large spheres.
In the AO model, customarily the $\eta_{\rm l}$--$\mu_{\rm s}$ plane is chosen but the topology of phase diagrams
is very similar to the one in the $\eta_{\rm l}$--$P$ plane. 

\subsubsection{Small size ratios $q$}

For a size ratio $q=0.15$, the phase diagram is shown in Fig.~\ref{fig:phase_q_0.15} in two different projections. 
For both HD and AO mixtures, the addition of the small species leads to an increased coexistence pressure for
the fluid--crystal transition, i.e. the fluid phase is stabilized. The AO mixture shows the typical widening of the
coexistence gap ($\eta_{\rm l,cr}-\eta_{\rm l,fl}$) with increasing concentration of the small species
(see Fig.~\ref{fig:phase_q_0.15_etal}), smoothly leading to a sublimation line. 
For $\eta_{\rm s} \lesssim 0.01$, the HD mixture binodal follows the AO binodal, i.e.
also shows an initial widening of the coexistence gap. This could be expected since for these small concentrations
the small disks only act as depletants and their mutual interaction is irrelevant. For higher $\eta_{\rm s}$, the 
binodals separate. The choice of the parameter $a$ in the functional has a significant influence on the location of the binodal.
This is similar to the observation in Ref.~\cite{mortazavifar2016fundamental} that also in the 3D case, the
binodal differs considerably between the White Bear II (tensor) and the Rosenfeld (tensor) functional, although the
differences in the one--component case are not that significant.     

For the size ratios $q=0.3$ and $q=0.45$, the phase diagrams are shown in Fig.~\ref{fig:phase_diagrams_smallq} in the
$\eta_{\rm l}$--$P$ plane. For the AO mixture, the liquid (rich in large disks)--vapor (poor in large disks) transition
has become stable which leads to the appearance of a triple point above which sublimation (vapor--crystal transition) is
stable. The triple point pressure decreases with increasing $q$. 
The difference in the location of the  liquid--vapor transition between
the FMT results for the two different values of $a$ is only a consequence of normalizing the pressure axis by
$P_{\rm co}^{\rm 1c}$ (for the two $a$ values, it differs by $\sim$ 15\%, see Table \ref{tab:co_HD}).
For the HD mixtures, there is no fluid--fluid transition and there is hardly any widening of the coexistence gap
of the fluid--crystal transition visible.

The results for the AO mixture are very similar to the 3D case \cite{mortazavifar2016fundamental}. Experimentally, it is
possible to realize such 2D systems by sedimented monolayers of colloidal spheres 
(as in Refs.~\cite{Thorneywork2014,thorneywork2017two}) to which nonadsorbing polymers can be added.  
For small size ratios $q \lesssim 0.15$ it would be interesting to study experimentally or by simulations 
the fate of the established melting scenario for hard disks as the polymer concentration is increased. As we have
seen, the coexistence gap continuously widens in this case, and we expect that towards the sublimation regime only
the first--order transition survives. 

\subsubsection{Size ratios $q$ close to 1}

For size ratios $q$ in the vicinity of 1, we only focus on the HD mixture. In the AO mixture, the phase diagram
becomes rather uninteresting with regard to crystal phases. There, upon addition of the smaller, polymeric
component the one--component crystal does not change very much: the polymers fill up the vacancies until the triple point is 
reached and the fluid--crystal transition becomes unstable with respect to sublimation. Again this is very similar to the 3D
case, and a detailed discussion can be found in Ref.~\cite{mortazavifar2016fundamental}.
  
For HD mixtures, phase diagrams are shown in Fig.~\ref{fig:phase_diagram_big_q}. For $q$ very close to 1, the 
phase diagram is of a type commonly denoted as spindle type (which would be directly visible in the $c_{\rm l}$-$P$ plane
or in the $c_{\rm s}$-$P$ plane): 
The coexistence pressure 
continuously increasing upon addition of smaller disks and reaches its maximum for the pure small-disk system
(see Fig.~\ref{fig:phase_diagram_big_q}(a))
Upon lowering $q$, the type of phase diagram crosses over to azeotropic (see Fig.~\ref{fig:phase_diagram_big_q}(b) 
and (c)): there, a maximum pressure for a stable fluid is found for a certain finite composition, i.e. for a truly
mixed system. At this point of maximum pressure, the coexisting fluid and crystal have the same composition (azeotropic point). 
The precise value for $q$ where this transition happens depends on the parameter $a$ in the functional;
it is around 0.91  for $a=3$ and around 0.93 for $a=11/4$.  The transition from spindle--type to azeotropic phase diagrams has also been observed in simulations of hard sphere mixtures in 3D \cite{Frenkel1991}. There, the transition happens at around $q=0.94$. Furthermore, in 3D the azeotropic phase diagram changes to a eutectic phase diagram already at around $q=0.88$. From our results, this happens in 2D at 
much lower $q$ (see below).

\subsubsection{Intermediate size ratios $q$}

Again we will only discuss HD mixtures. The phase diagram
for $q=0.6$  is shown in Fig.~\ref{fig:phase_diagram_intermediate_q}(a) and for $q=0.7$ in 
Fig.~\ref{fig:phase_diagram_intermediate_q}(b),(c). For $q=0.6$, we observe a phase diagram of eutectic type. 
It is actually very similar to the phase diagram found in simulations for $q=1/1.4$ (see Ref.~\cite{russo2017disappearance}
Supplementary Material). The crossover to the azeotropic phase diagram (as seen in 
$q=0.75$ in Fig.~\ref{fig:phase_diagram_big_q}(c)) is surprising according to the FMT results.
 
For $q=0.7$, a 3 dimensional phase diagram in $\eta_{\rm l}$--$\eta_{\rm s}$--$P$ space is presented in Fig.~\ref{fig:phase_diagram_intermediate_q}(c). The coexisting plane with a majority of large disks (black surface) is close to the one of small disks (blue surface), but does not cross. By increasing $q$, two plane touch then form an azeotropic type.  Back to the $\eta_{\rm l}$--$P$ plane, the branch with a majority of large disks distorts to form an azeotropic point (see the black lines in
Fig.~\ref{fig:phase_diagram_intermediate_q}(b)) whereas the branch with a majority of small disks remains approximately
unchanged when compared with $q=0.6$ (blue lines in Fig.~\ref{fig:phase_diagram_intermediate_q}(b)).  Thus, above the azeotropic point pressure there is a stable and a metastable coexistence between a crystal with a majority of small disks
and a mixed fluid. 

\begin{figure}[t]
    \begin{subfigure}[bl]{0.5\textwidth}
        \includegraphics[width=\textwidth]{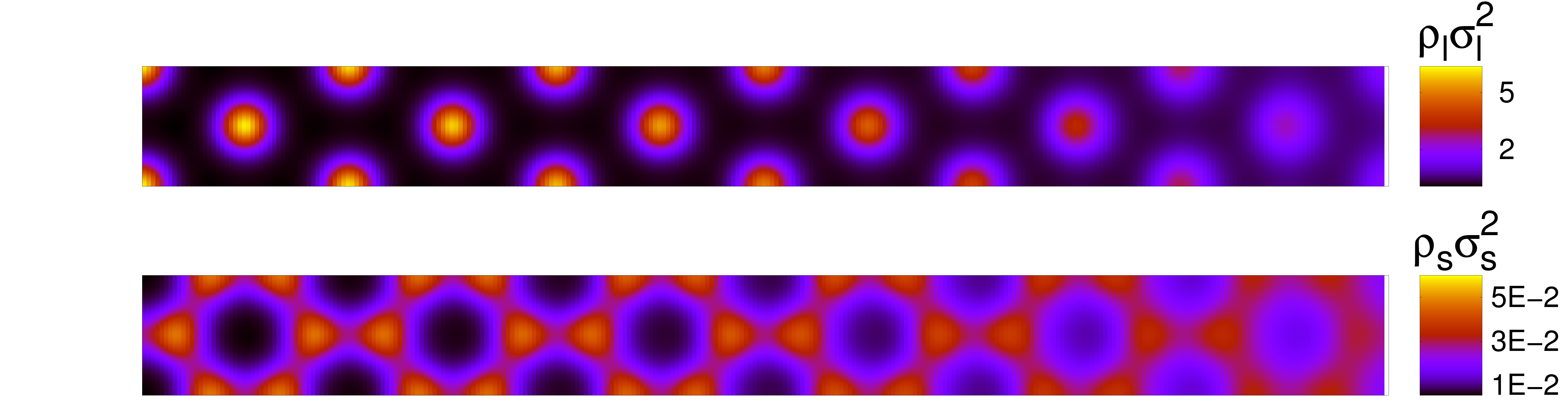}        
        \subcaption{$q=0.15,\cs=0.50$}
        \label{fig:example_q_0.15}
    \end{subfigure}
    \begin{subfigure}[bl]{0.5\textwidth}
        \includegraphics[width=\textwidth]{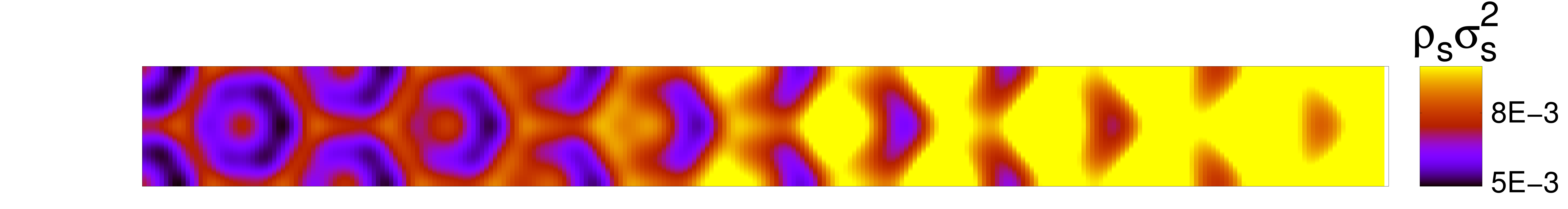}
        \subcaption{$q=0.45,\cs=0.03$}
        \label{fig:example_q_0.45_s_low}
    \end{subfigure}
    \begin{subfigure}[bl]{0.5\textwidth}
        \includegraphics[width=\textwidth]{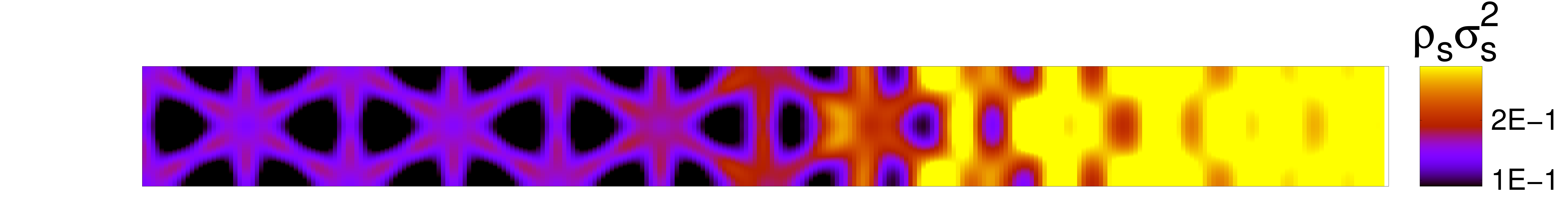}
        \subcaption{$q=0.45,\cs=0.39$}
        \label{fig:example_q_0.45_s}
    \end{subfigure}
    \begin{subfigure}[bl]{0.5\textwidth}
        \includegraphics[width=\textwidth]{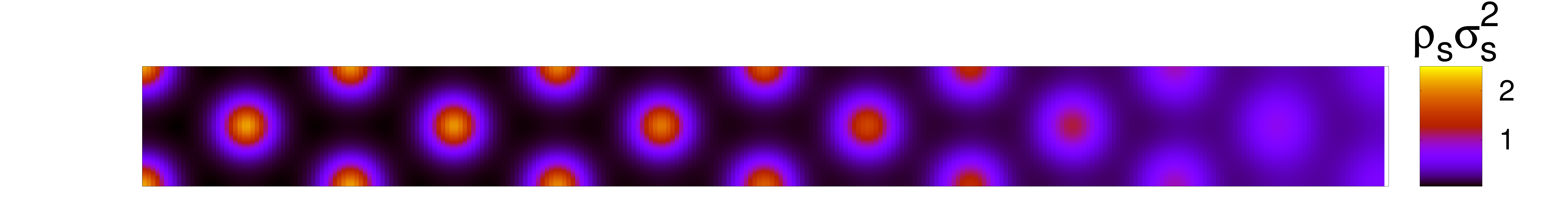}
        \subcaption{$q=0.75,\cs=0.52$}
        \label{fig:example_q_0.75_s}
    \end{subfigure}
    \caption{HD mixture density profiles $\rho$ cross crystal(left)-fluid(right) interface for $q=0.15,0.45$ and $0.75$. Since all large disks density profiles looks similar, here we only show a representative one.}
    \label{fig:example_interface}
\end{figure}    
\begin{figure*}[t]
    \begin{subfigure}[bl]{\textwidth}
        \includegraphics[width=\textwidth]{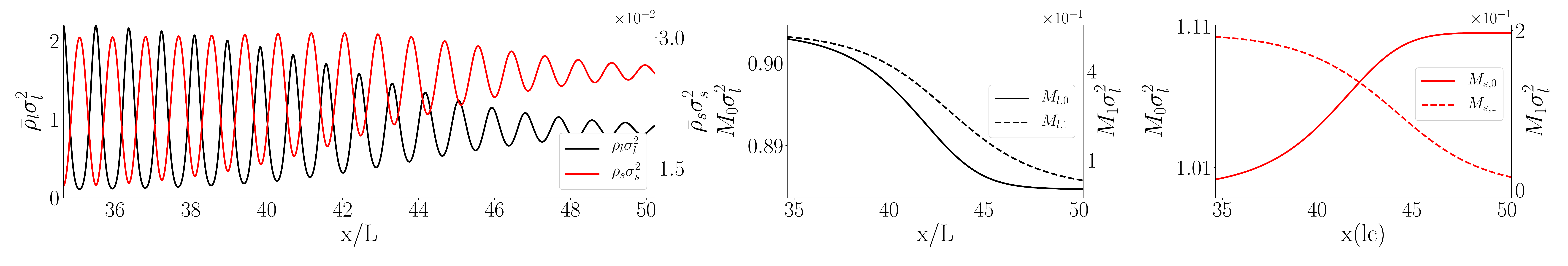}
        \subcaption{$q=0.15,\cs=0.50$}
        \label{fig:mode_q_0.15}
    \end{subfigure}
    \begin{subfigure}[bl]{\textwidth}
        \includegraphics[width=\textwidth]{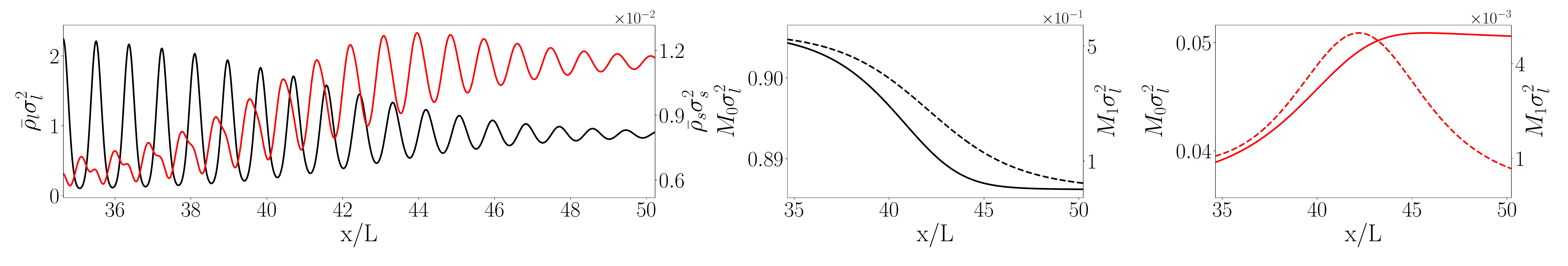}
        \subcaption{$q=0.45,\cs=0.03$}
        \label{fig:mode_q_0.45_low}
    \end{subfigure}
        \begin{subfigure}[bl]{\textwidth}
        \includegraphics[width=\textwidth]{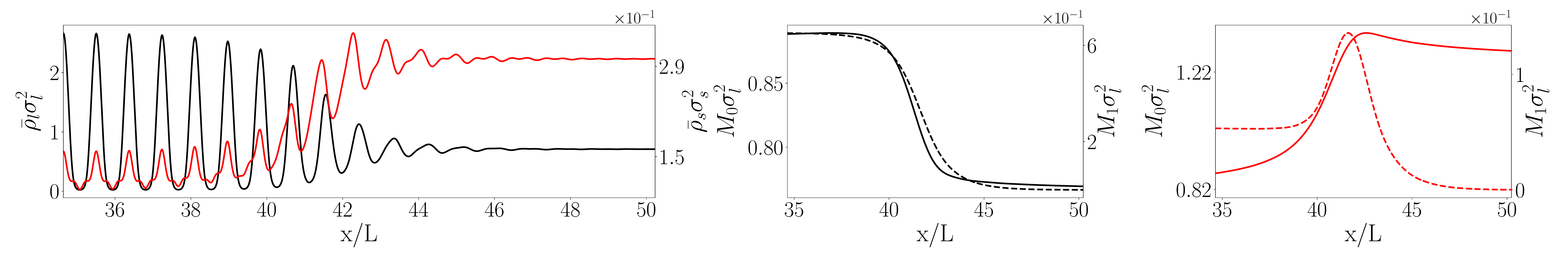}
        \subcaption{$q=0.45,\cs=0.39$}
        \label{fig:mode_q_0.45}
    \end{subfigure}
        \begin{subfigure}[bl]{\textwidth}
        \includegraphics[width=\textwidth]{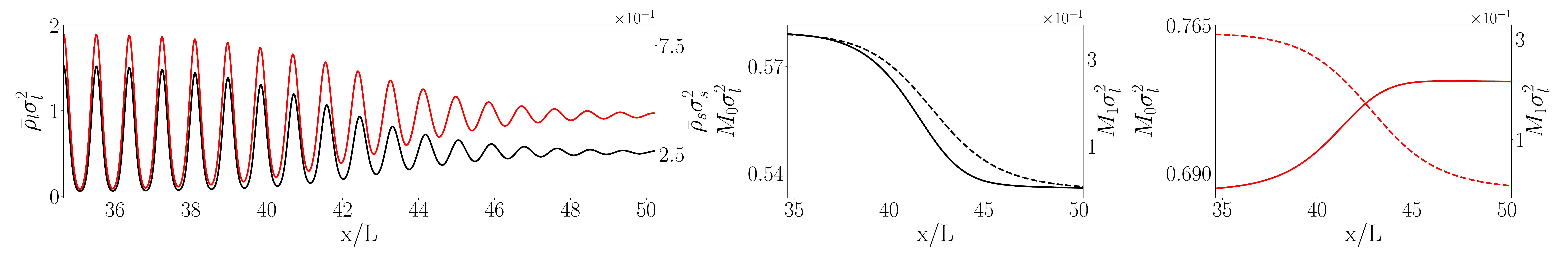}
        \subcaption{$q=0.75,\cs=0.52$}
        \label{fig:mode_q_0.75}
    \end{subfigure}
    \caption{Laterally averaged density profiles (left column), density and leading crystallinity modes of large disks (middle column)
and small disks (right column) for the four interfaces of Fig.~\ref{fig:example_interface}. Black lines refer to large disks, red lines
to small disks. In the middle and right column, full lines are density modes $M_0$ and dashed lines crystallinity modes $M_1$.} 
    \label{fig:mode_expand}
\end{figure*} 
\begin{figure*}[t]
    \begin{subfigure}[b]{0.4\textwidth}
        \includegraphics[width=\textwidth]{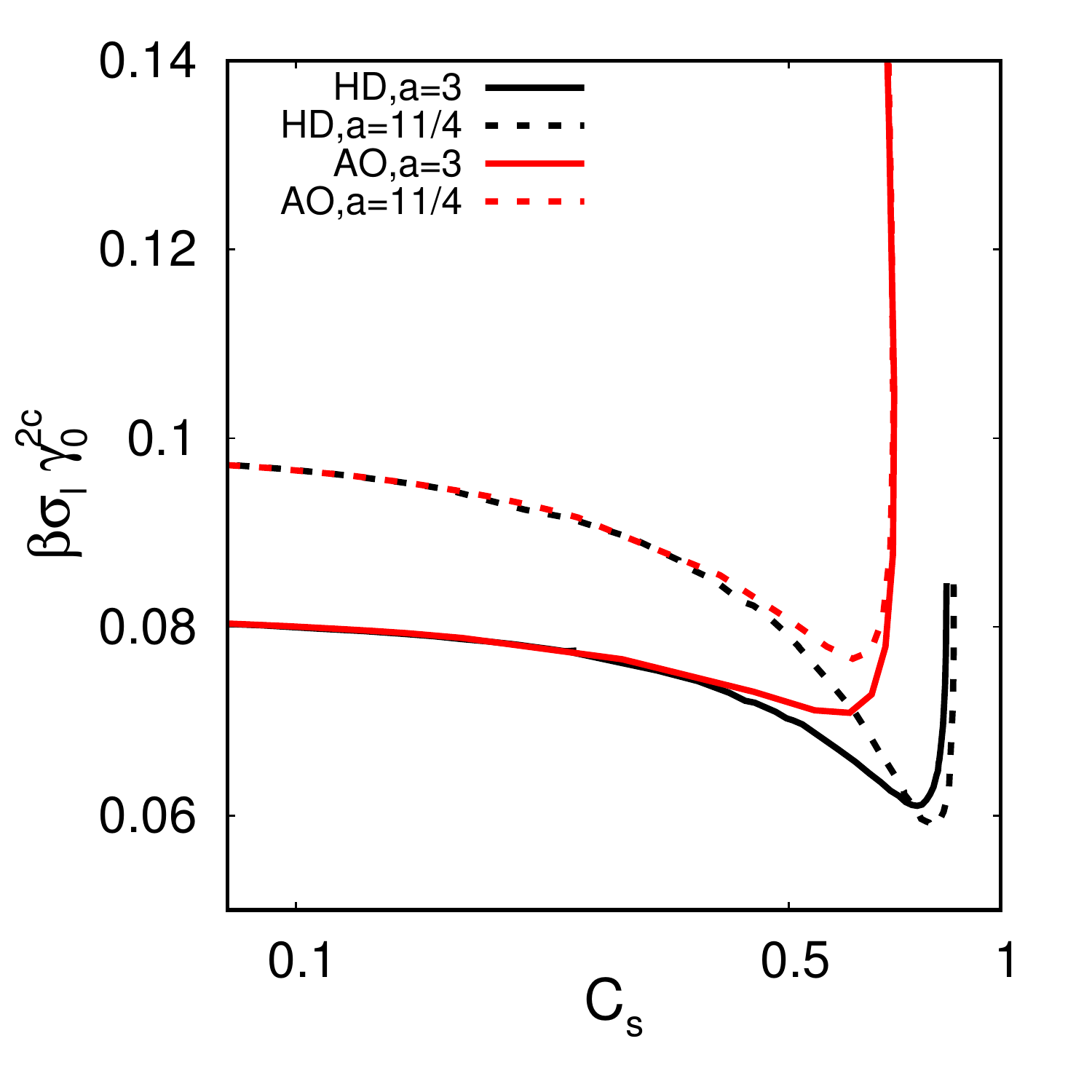}
        \caption{$q=0.15$}
        \label{fig:gamma_0.15}
    \end{subfigure}
        \begin{subfigure}[b]{0.4\textwidth}
        \includegraphics[width=\textwidth]{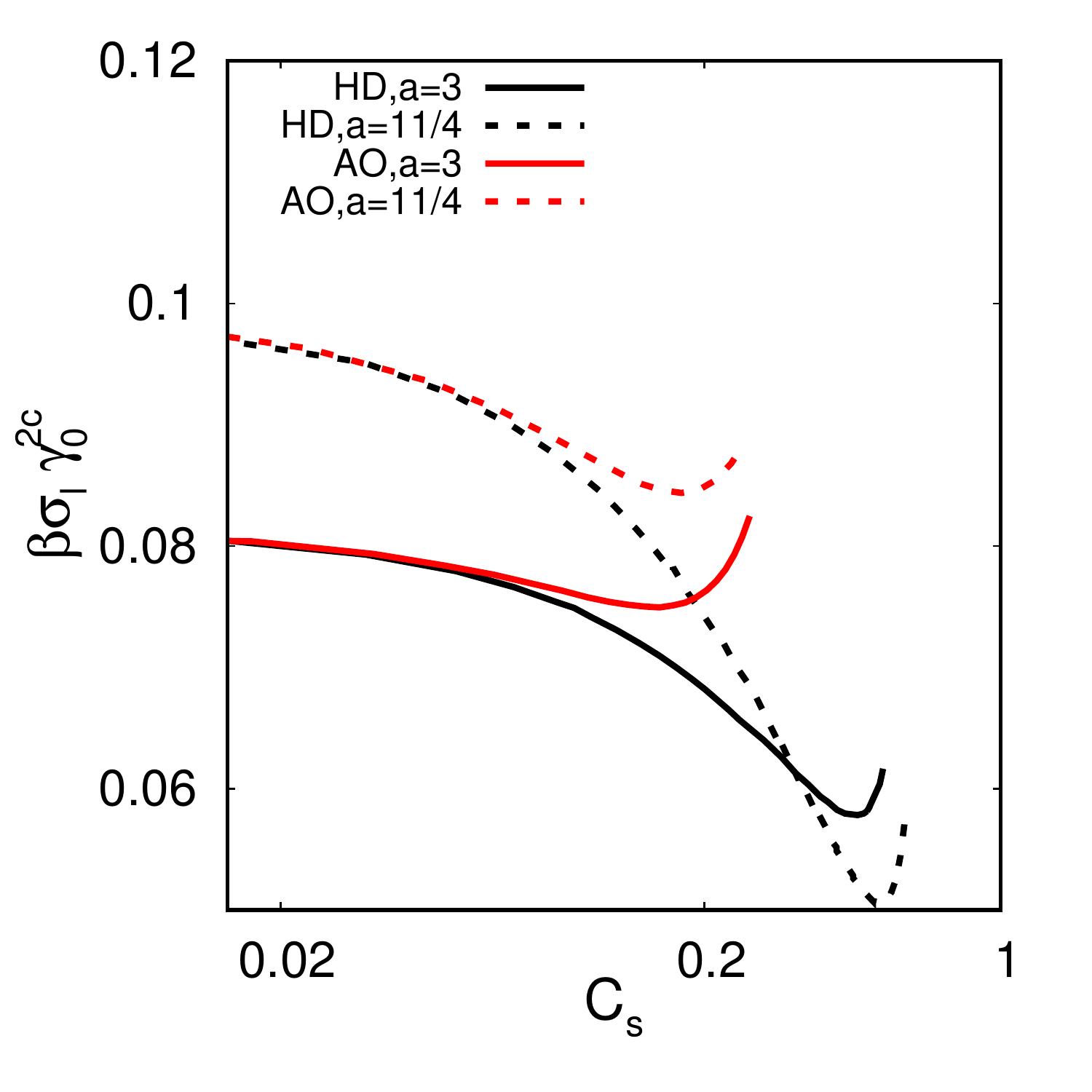}
        \caption{$q=0.3$}
        \label{fig:gamma_0.3}
    \end{subfigure}
        \begin{subfigure}[b]{0.4\textwidth}
        \includegraphics[width=\textwidth]{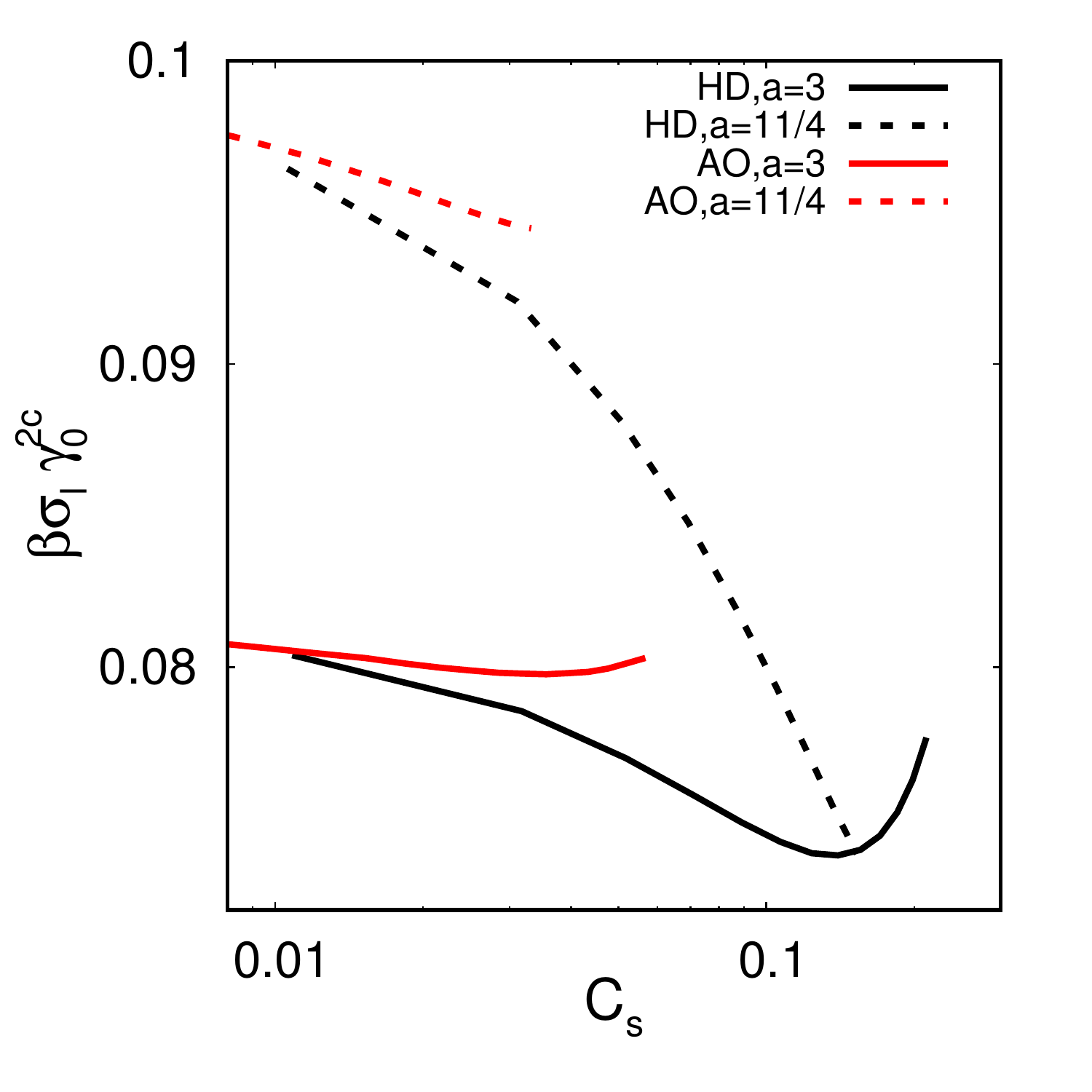}
        \caption{$q=0.45$}
        \label{fig:gamma_0.45}
    \end{subfigure}
    \begin{subfigure}[b]{0.4\textwidth}
        \includegraphics[width=\textwidth]{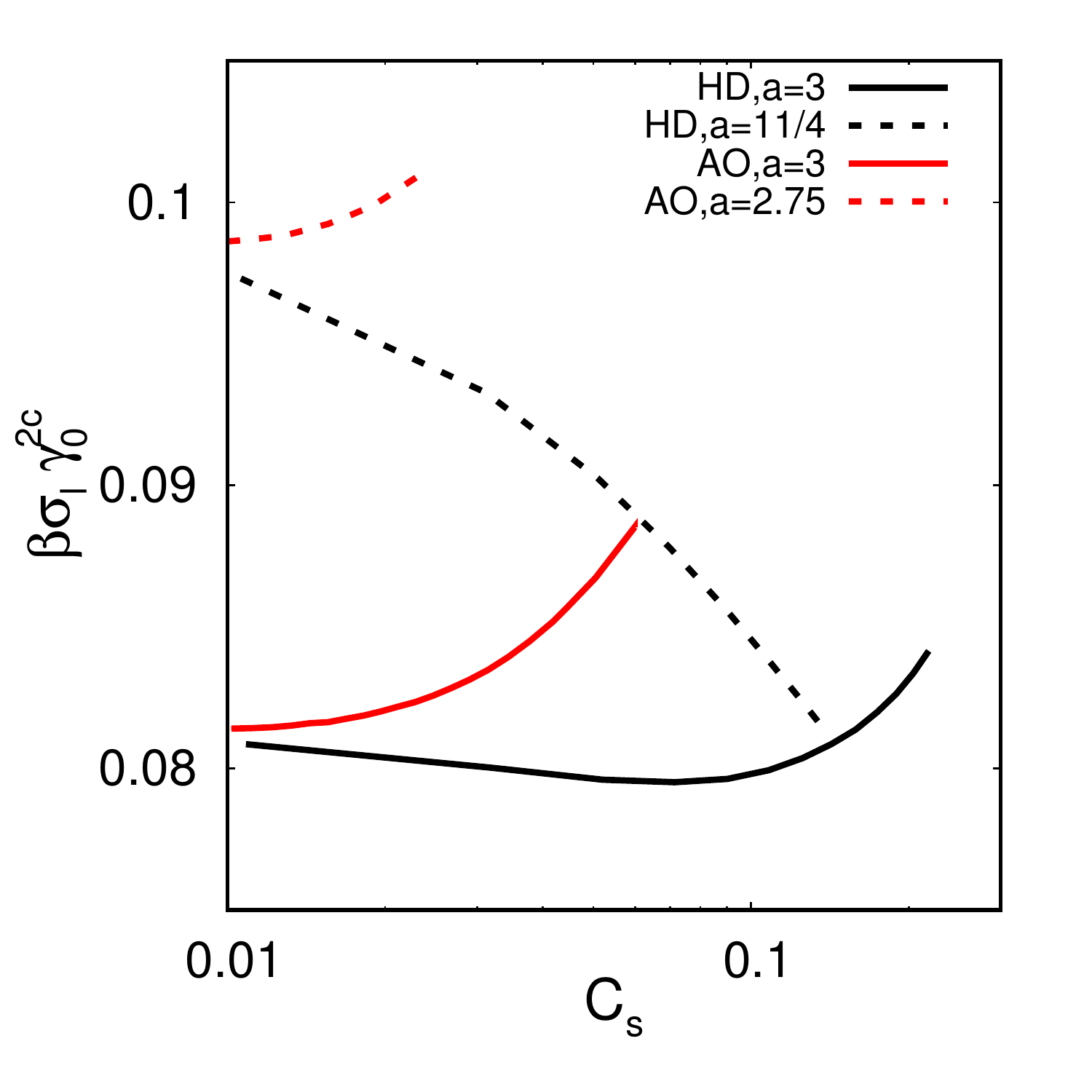}
        \caption{$q=0.6$}
        \label{fig:gamma_0.6}
    \end{subfigure}
    \caption{The crystal--fluid surface tension $\gamma^{\rm 2c}_0$ for both the AO and the HD case and for two values of $a$. 
Note that $\cs$ is in log scale. (\subref{fig:gamma_0.15}) Size ratio $q = 0.15$, (\subref{fig:gamma_0.3}) $q = 0.3$, (\subref{fig:gamma_0.45}) $q = 0.45$ and (\subref{fig:gamma_0.6})  $q = 0.6$.}
    \label{fig:gamma_cr_l}
\end{figure*} 

\begin{figure*}[t]
	\begin{subfigure}[b]{0.32\textwidth}
	\includegraphics[width=\textwidth]{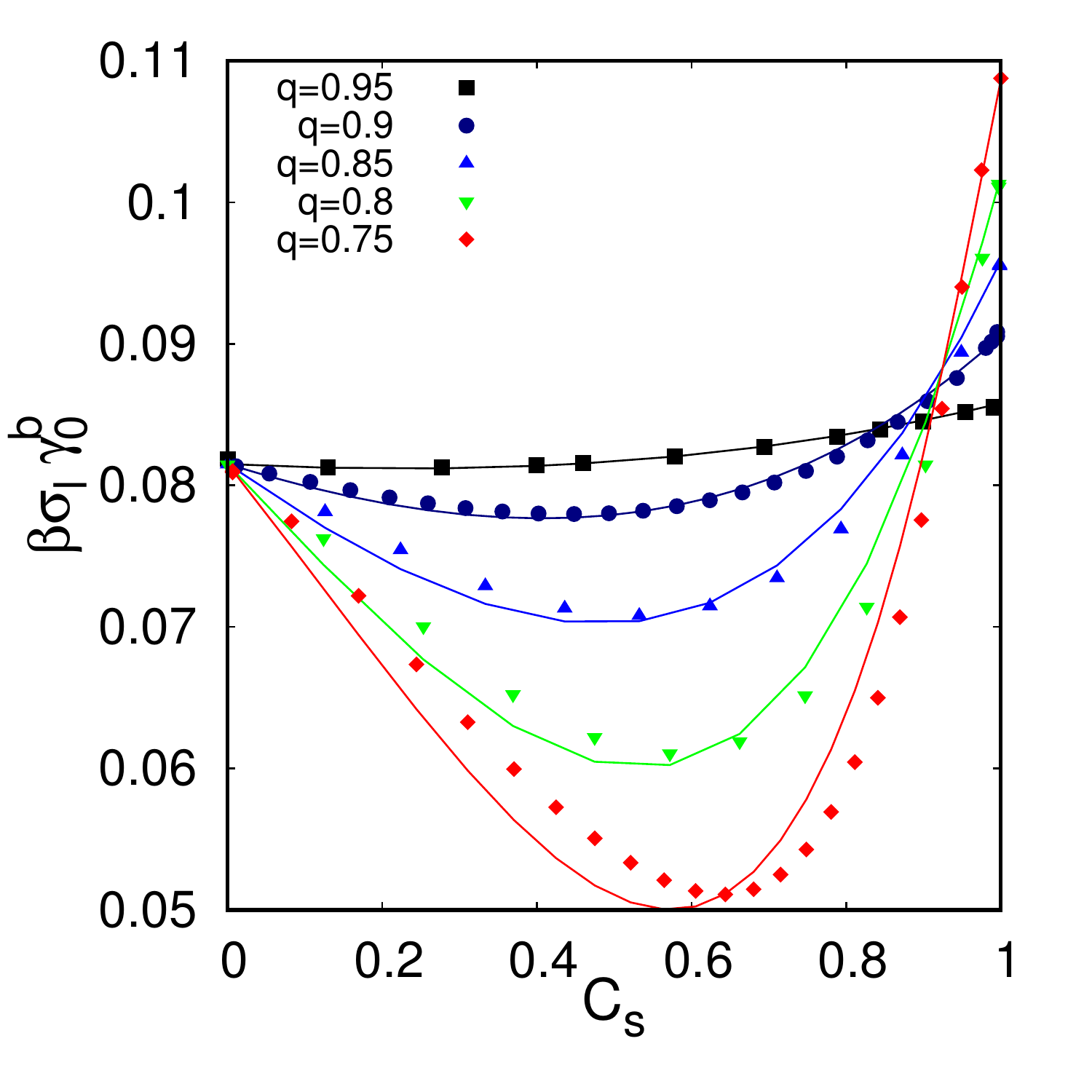}
	\caption{$a=3$}
    \label{fig:gamma_a_3_all}
    \end{subfigure}
	\begin{subfigure}[b]{0.32\textwidth}
	\includegraphics[width=\textwidth]{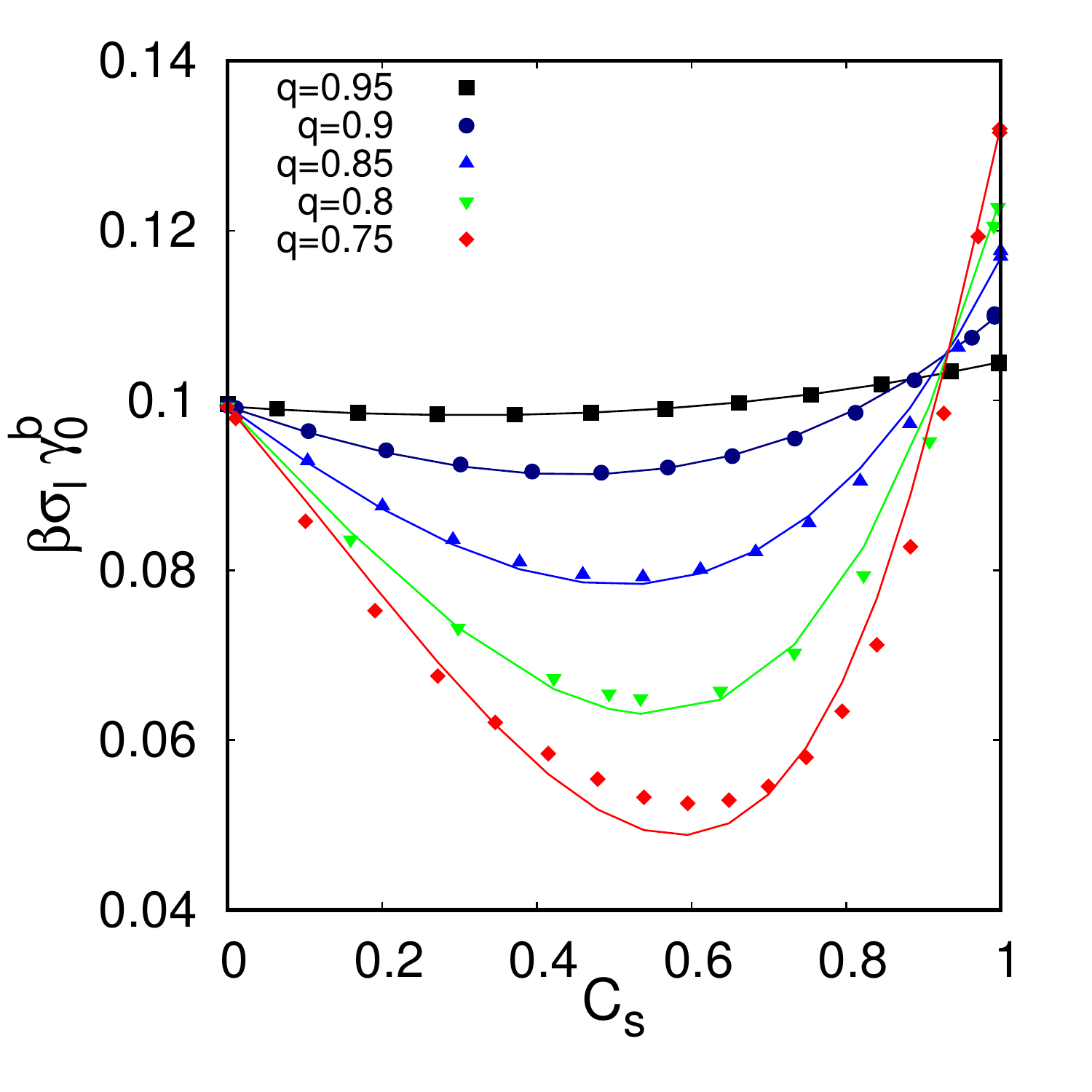}
	\caption{$a=11/4$}
    \label{fig:gamma_a_2.75_all}
    \end{subfigure}
    \centering{
    \begin{subfigure}[b]{0.32\textwidth}
	\includegraphics[width=\textwidth]{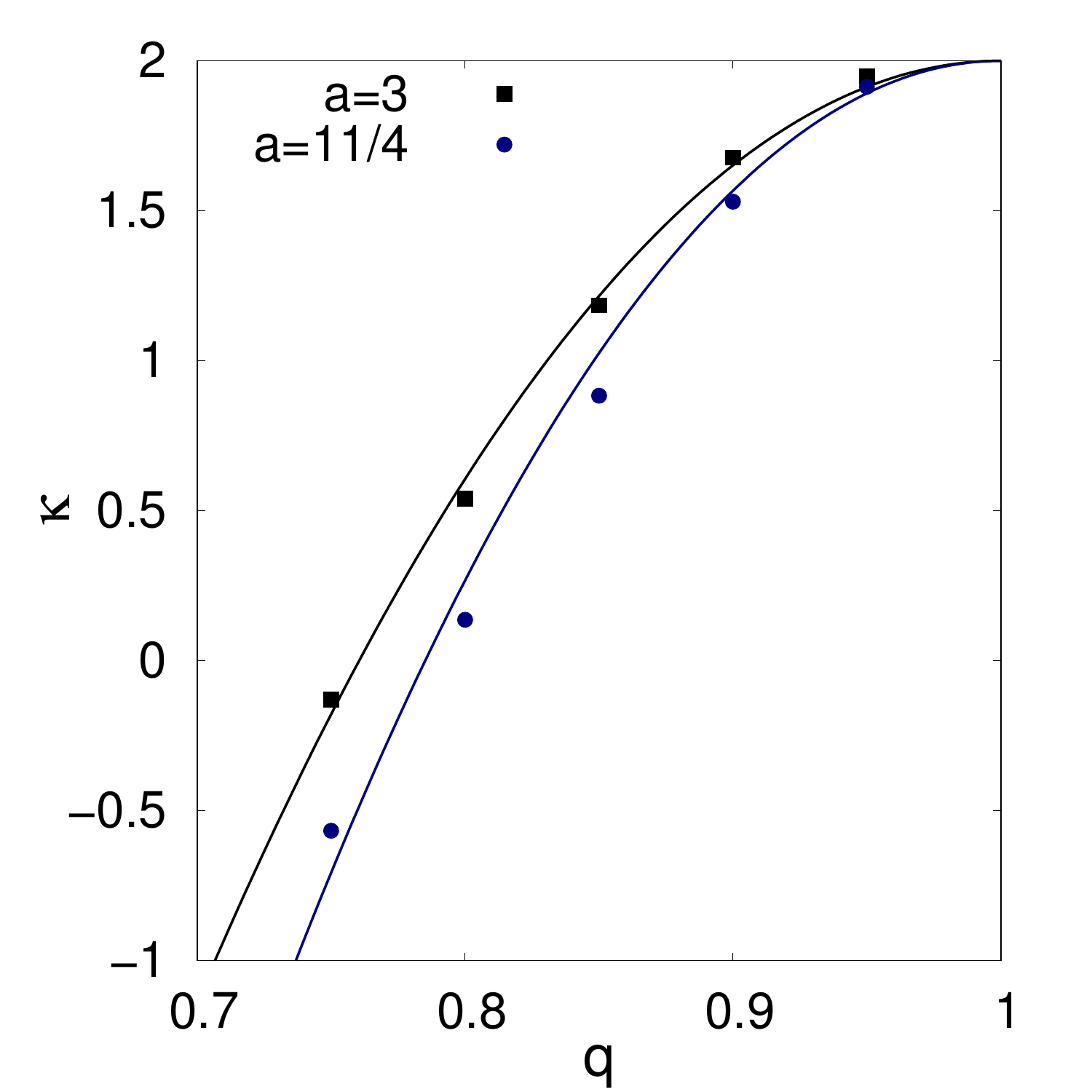}
	\caption{$\kappa$}
    \label{fig:a_q_fitting}
    \end{subfigure}
    }
\caption{The crystal--fluid surface tension $\gamma^{\rm 2c}_0$ for the HD case and for size ratios $q \ge 0.75$. 
The value of the parameter in the functional $a=3$ (\subref{fig:gamma_a_3_all}) and $a=11/4$ (\subref{fig:gamma_a_2.75_all}). 
Symbols are the numerical results and the lines are best fits to Eq.\eqref{eqn:fitting}.  
The fit values $\kappa(q)$ are shown in (\subref{fig:a_q_fitting}), lines are functions
 $\kappa=\kappa_1(q-1)^2+2$ with $\kappa_1=-34.86$ and $-43.32$ for $a=3$ and $11/4$, respectively.}
\label{fig:gamma_l_s_big}
\end{figure*}

\subsection{Binary systems: Interface density profiles}
Fiq.~\ref{fig:example_interface} shows representative density profiles of the crystal--fluid interface for hard disk mixtures with $q=0.15,0.45$ and $0.75$. 
For all size ratios $q$, the density of large disks is always peaked on the triangular lattice sites (see Fig.~\ref{fig:example_q_0.15}, upper panel)   
while the density of small disks changes from interstitial to substitutional disorder by increasing $q$ (see also the discussion in Sec.~\ref{sec:crystaldensity}). For the AO mixture, we found similar density profiles for $q <0.5$, except Fig.~\ref{fig:example_q_0.45_s}. From the profiles one infers a rather broad interface.

 We analyze the interface structure further by employing the methods of Ref.~\cite{oettel2012mode}. Smooth average density and crystallinity modes can be extracted from the Fourier transform of the full density profiles by picking a lateral reciprocal lattice vector ($\mathbf{K}_y$) and cutting out a window around a reciprocal lattice vector $\mathbf{K}_x$ parallel to the interface normal. The average modes are the inverse Fourier transforms of the cut--out window. The average density mode $M_0$ is obtained by choosing $\mathbf{K}_x=\mathbf{K}_y=0$ and the leading crystallinity mode $M_1$ is obtained by choosing $\mathbf{K}_y=0, \mathbf{K}_x= 4\pi/\left(\sqrt{3}L\right)$ where $L$ is the length of the rectangular unit cell side which is parallel to the interface, see Figs.~\ref{fig:example_unit_cell} and \ref{fig:example_interface}. $M_1$ is complex in general, in figures we show its absolute value only. 

In Fig.~\ref{fig:mode_expand} we compare laterally averaged density profiles with the extracted density and crystallinity modes
for the four interfaces of Fig.~\ref{fig:example_interface}. Several observations can be made. First, looking at 
the density and crystallinity mode of large disks (middle column in Fig.~\ref{fig:mode_expand}) we note that coming from the
fluid side, crystallinity sets in earlier as densification (except for the case $q=0.45$, $\cs=0.39$). This
has also been noted before in the 3D case of one--component hard spheres \cite{oettel2012mode}. Second, looking at the
 density and crystallinity mode of small disks (right column in Fig.~\ref{fig:mode_expand}) we observe that for small
$q=0.15$ (interstitial disorder) and large $q=0.75$ (substitutional disorder) the small disk crystallinity is essentially
proportional to the large disk crystallinity. Since the crystal has a smaller concentration of small disks than in the fluid, the density
mode increases from left to right but stays monotonic. For the intermediate size ratio $q=0.45$ we note that the crystallinity
of small disks is peaked at the interface, and for $\cs=0.39$ this also applies to the density mode. Thus we see an interfacial enrichment of ordered, small spheres. This interfacial enrichment can be also seen in the laterally averaged density
profiles  (left column in Fig.~\ref{fig:mode_expand}) which exhibit an increase in the oscillation amplitude of the
small sphere density (red lines) in the interfacial region. However, the quantification of this effect is easier using the
crystallinity and density modes.

\subsection{Binary systems: Crystal--fluid surface tensions}

\subsubsection{Size ratio $q \le 0.6$}
For small to moderate size ratios of up to 0.6, we may view the small disks as depletants, at least for 
small concentrations $\cs$. In Fig.~\ref{fig:gamma_cr_l}, we show the associated crystal-fluid planar surface tension $\gamma_0^{\rm 2c}$ 
versus $\cs$ for both AO and HD mixtures.
 
For $q=0.15$ (Fig.~\ref{fig:gamma_0.15}), we have computed the surface tension for  $\cs$ up to 1.
We remind the reader of the associated phase diagrams (Fig.~\ref{fig:phase_q_0.15}) which in the AO case
shows the typical widening of the coexistence gap. In the HD case, the widening of the coexistence gap
follows the AO case only for small $\cs$. It is a bit surprising that the 
surface tension {\em decreases} upon addition of small disks, with the results for the HD mixture are on top
of those for the AO mixture until $\cs \sim 0.4$. In the depletion picture, the addition of small disks leads
to an increasing attraction between large disks. In mean--field approximation, the increasing attraction together
with an increasing coexistence gap should lead to a higher surface tension. Such an increase is seen both for the AO model
and the HD case only for rather large $\cs$, after a minimum has been reached around $\cs \approx 0.6$ (Fig.\ref{fig:gamma_0.15}).
In the HD case, for $\cs \to 1$ we reach the monocomponent case for small disks, thus the surface tension should reach
$\gamma_0^{\rm 2c}(\cs=1) = \frac{\sigl}{\sigs}\gamma_0^{\rm 2c}(\cs=0) = \gamma^{\rm 1c}_0/q$. 

The peculiar behavior of an initially decreasing surface tension is also seen for $q=0.3$ (Fig.~\ref{fig:gamma_0.3}),
$q=0.45$ (Fig.~\ref{fig:gamma_0.45}) and $q=0.6$ (Fig.~\ref{fig:gamma_0.6}), although the decrease becomes smaller
with increasing $q$. With increasing size ratio, also the HD and the AO results differ more and more already for small
$\cs$ and we also note that the choice of the parameter $a$ in the FMT functional influences the results considerably.  Overall, the surface tensions are rather small on the thermal energy scale. For the monocomponent case this
leads to strong interface fluctuations, as observed in Ref.~\cite{thorneywork2017two}. Owing to the decrease
in $\gamma^{\rm 2c}_0$ upon addition of small disks, we would expect that these fluctuations also become stronger.  

\subsubsection{Size ratio $q \ge 0.75$: HD mixtures}

For $q\geq 0.75$, the phase diagram in the HD mixture is of azeotropic or spindle type (see Fig.~\ref{fig:phase_diagram_big_q}), 
thus we can determine $\gamma_0^{\rm 2c}$ in the whole range of concentrations from $\cs=0$ up to $1$. 
In Fig.\ref{fig:gamma_l_s_big}, the surface tension $\gamma_0^{\rm 2c}$ versus $\cs$ is shown for four aspect rations
$q\geq 0.75$ and the two values of the parameter $a$. Qualitatively, there is no significant dependence on $a$
for these size ratios. As before (for small $q$) the initial decrease of $\gamma_0^{\rm 2c}$
for small $\cs$ is present. There is a minimum in the surface tension around $\cs=0.5$ and it reaches the
correct monocomponent value $\gamma_0^{\rm 2c}(\cs=1) = \gamma_0^{\rm 1c}/q$.

The surface tensions can actually be well described with the following function involving one fit parameter $\kappa$:
\begin{equation}
\label{eqn:fitting}
  \gamma_0^{\rm 2c}(\left[\eta_{\rm cr}\right],q)=\frac{\gamma_0^{\rm 1c}}{(\eta^{\rm 1c}_{\rm cr})^2}
     \left(({\eta_{\rm l,cr}})^2+\frac{({\eta_{\rm s,cr}})^2}{q}+\kappa \eta_{\rm l,cr}\,\eta_{\rm s,cr} \right),
\end{equation}
where $\gamma_0^{\rm 1c}$ and $\eta^{\rm 1c}_{\rm cr}$ on the right hand side of Eq.\eqref{eqn:fitting} are 
the monocomponent surface tension and the coexistence crystal packing fraction (see Table \ref{tab:co_HD}) and 
$\eta_{\rm l/s,cr}$ are the coexistence crystal packing fractions of large/small HD.
For the fit parameter $\kappa$ we note that 
$\lim_{q\to 1}\kappa(q)=2$. For $q<0.75$, Eq.~\eqref{eqn:fitting} is not valid, which may be due to the 
complicated transition from an azeotropic to an eutectic phase diagram (as discussed before).
 
\section{Summary and conclusion}

Using density functional theory (fundamental measure theory),
we have performed an extensive study of the phase diagram and crystal--fluid surface tensions in binary hard disk systems,
both for the additive case and the non--additive (Asakura--Oosawa like) case. Since we assumed a periodic crystal, 
we find first--order transitions only. These correspond to the first--order fluid--hexatic transition for the 
one--component case and presumably to first--order fluid--crystal transitions (which become stable upon admixing a second 
component, see e.g. Ref.~\cite{russo2017disappearance}). Overall, the phase diagrams are qualitatively very similar
to 3D. In the AO case and for small size ratios $q$, the typical continuous widening of the coexistence gap is observed
upon addition of the smaller species, and for intermediate $q$ a vapor--liquid transition becomes stable.
In the additive case, the phase diagrams show the sequence spindle $\to$ azeotropic $\to$ eutectic upon lowering $q$ from 
1 to 0.6 (similar to 3D). However, the transition from azeotropic to eutectic is different from what is known in
3D hard sphere systems (see the phase diagram in Fig.~\ref{fig:phase_diagram_intermediate_q}(b),(c) for $q=0.7$).
 
The results for the crystal-fluid surface tensions reveal two things. Overall, their values are much smaller than 1
in thermal units $1/(\beta \sigma_{\rm l})$. For the one--component case, the resulting large thermal fluctuations of the interface
have been observed experimentally \cite{thorneywork2017two}. Secondly, the addition of a second component leads in general
to a substantial decrease in the surface tension. This holds for the AO case (for $q \lesssim 0.6$) and also for the
additive case (here for the whole range of $q$). Complementary, dedicated simulation or experimental results on this
are clearly desirable, also in view of the relevance of the surface tension for  nucleation processes, see
Ref.~\cite{Gonzalez2016} for a review on more qualitative results on 2D crystal and defect formation. The observed
decrease in surface tension should lead to a considerable decrease in the time scales of crystal nucleation.
    
In contrast to phase diagrams, results on crystal--fluid surface tensions in binary 3D systems are scarce.
For binary hard spheres with a size ratio of $q=0.9$, results are reported in Ref.~\cite{Laird2008}. For  
that $q$, the phase diagram is azeotropic. The surface tension is found to increase monotonically with the 
addition of small spheres. These findings are similar to those for a 3D binary Lennard--Jones system
with zero size mismatch, but a ratio of interaction strengths of 0.75
(leading to a spindle--type phase diagram)  \cite{Foiles2009}, 
but they are different from the non-monotonic behavior found here in the 2D system
(see Fig.~\ref{fig:gamma_l_s_big}(a),(b)). 

The full minimization of the FMT functionals show interesting effects for the density distributions in the crystal unit cells
and of the crystal--fluid interfaces. For intermediate size ratios (examples shown for $q=0.45$) superpositions of substitutional and
alloy structures are found, and enhanced crystallinity and density of small disks is observed right at interface between crystal and fluid. Clearly, an
extension of the present studies to the global stability of alloy phases and their interfaces is desirable but requires
considerable more efforts.  
      
{\bf Acknowledgments:} S.--C. Lin thanks Landesgraduiertenf\"orderung Baden–
W\"urttemberg for financial support.
The authors acknowledge support by the High Performance and Cloud
Computing Group at the Zentrum f\"ur Datenverarbeitung of the University of T\"ubingen, the state of
Baden-W\"urttemberg through bwHPC and the German Research Foundation (DFG) through grant no INST 37/935-1
FUGG.

\appendix
\counterwithin{figure}{section}
\section{Liquid--vapor surface tension}
\begin{figure}[t]
    \begin{subfigure}[c]{0.23\textwidth}
        \includegraphics[width=\textwidth]{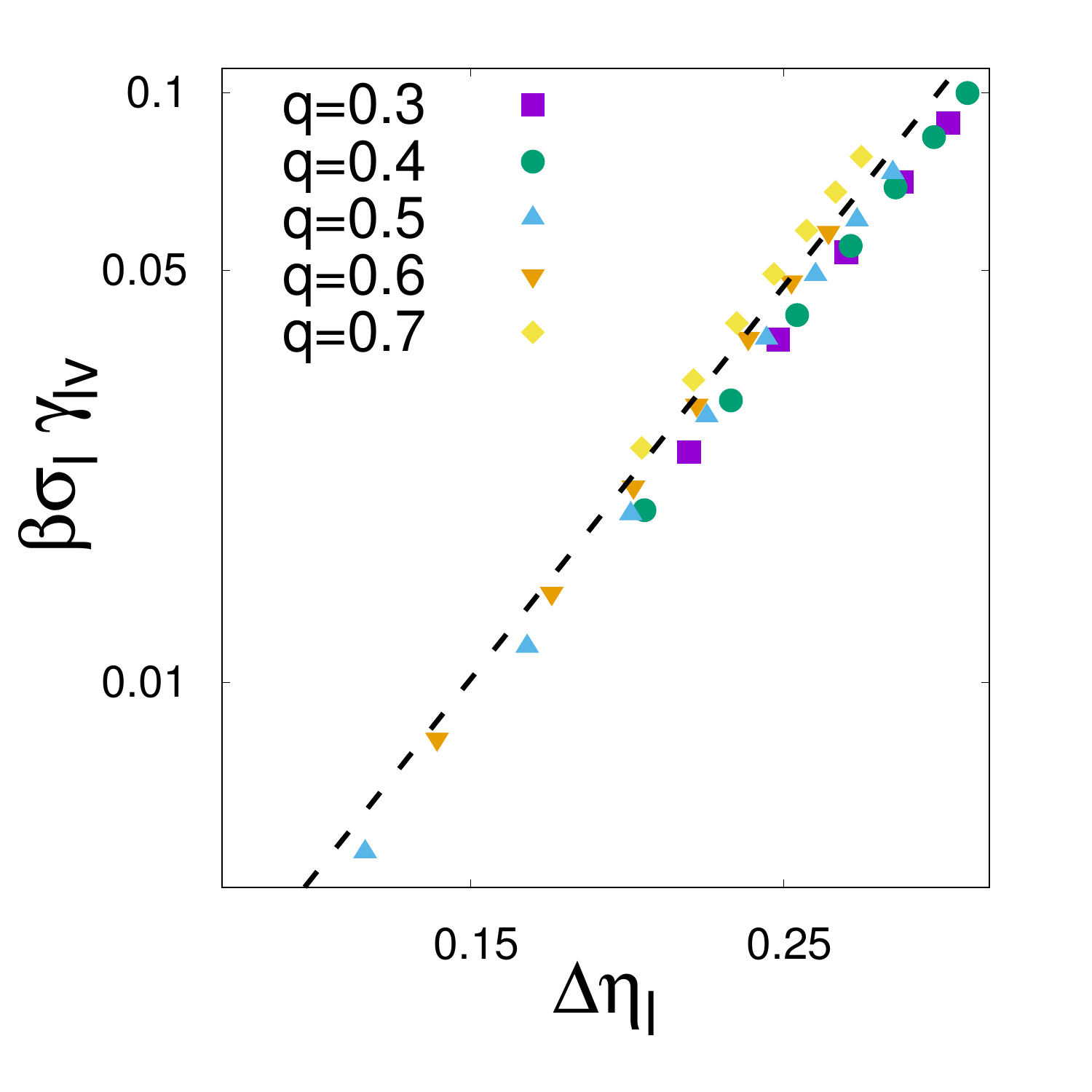}
        \caption{}
        \label{fig:gamma_lg}
    \end{subfigure}
    \begin{subfigure}[c]{0.23\textwidth}
        \includegraphics[width=\textwidth]{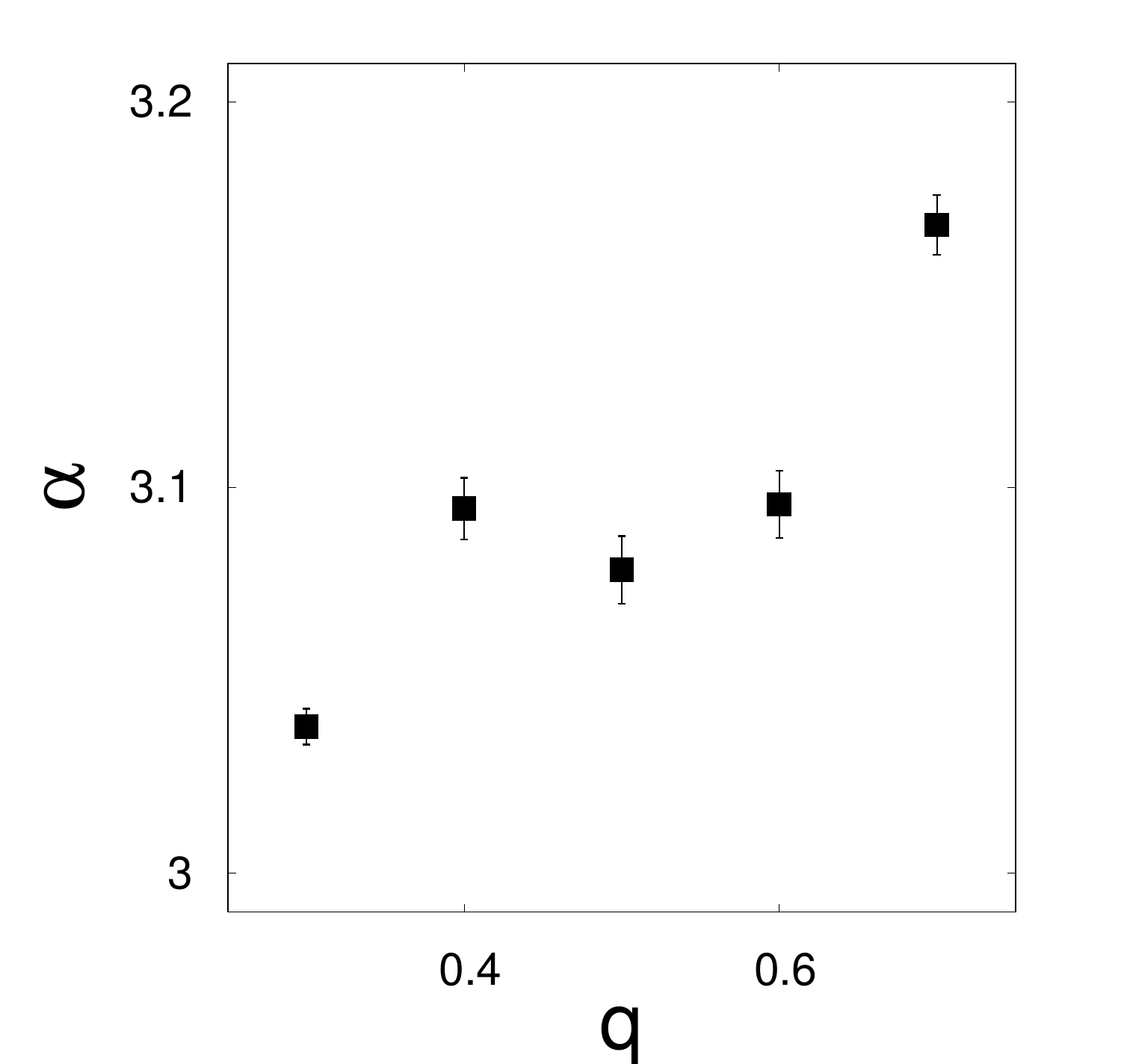}
        \caption{}
        \label{fig:crtical_exp}
    \end{subfigure}
    \caption{(\subref{fig:gamma_lg}) Double logarithmic plot of $\gamma_{\rm lv}$ versus $\Delta\eta_{\rm l}$. 
   The dashed line shows the mean--field proportionality  $\gamma_{\rm lv}\propto(\Delta\eta_{\rm l})^3$. (\subref{fig:crtical_exp}) Exponent $\alpha$ versus $q$, as extracted from linear fits to the results shown in (a).}
    \label{fig:gamma_liquid_gas}
\end{figure}

For completeness, we present also results for the liquid--vapor surface tension  $\gamma_{\rm lv}$ in the AO model for
size ratios $q=0.3...0.7$, see Fig.~\ref{fig:gamma_liquid_gas}(a). Similar to the crystal--fluid surface tension, the numerical
values for $\gamma_{\rm lv}$ are much smaller than 1
in thermal units $1/(\beta \sigma_{\rm l})$, even far away from the critical point. In Fig.~\ref{fig:gamma_liquid_gas}(b) we 
show the extracted exponent $\alpha$ for the assumed power--law relation $\gamma_{\rm lv} \propto \Delta \eta_{\rm l}^\alpha$,
where $\Delta \eta_{\rm l} = \eta_{\rm l,liq}-\eta_{\rm l, vap}$ is the difference between the coexistence packing fractions
of large disks in the liquid and vapor phase. For mean--field models, $\alpha=3$ close to the critical point, and this behavior is found to hold not only in the immediate vicinity of the critical point. This is similar to results from density functional studies of the 3D AO model \cite{vink2011fluid,vink2004grand}.
\newpage


\begin{thebibliography}{35}%
\makeatletter
\providecommand \@ifxundefined [1]{%
 \@ifx{#1\undefined}
}%
\providecommand \@ifnum [1]{%
 \ifnum #1\expandafter \@firstoftwo
 \else \expandafter \@secondoftwo
 \fi
}%
\providecommand \@ifx [1]{%
 \ifx #1\expandafter \@firstoftwo
 \else \expandafter \@secondoftwo
 \fi
}%
\providecommand \natexlab [1]{#1}%
\providecommand \enquote  [1]{``#1''}%
\providecommand \bibnamefont  [1]{#1}%
\providecommand \bibfnamefont [1]{#1}%
\providecommand \citenamefont [1]{#1}%
\providecommand \href@noop [0]{\@secondoftwo}%
\providecommand \href [0]{\begingroup \@sanitize@url \@href}%
\providecommand \@href[1]{\@@startlink{#1}\@@href}%
\providecommand \@@href[1]{\endgroup#1\@@endlink}%
\providecommand \@sanitize@url [0]{\catcode `\\12\catcode `\$12\catcode
  `\&12\catcode `\#12\catcode `\^12\catcode `\_12\catcode `\%12\relax}%
\providecommand \@@startlink[1]{}%
\providecommand \@@endlink[0]{}%
\providecommand \url  [0]{\begingroup\@sanitize@url \@url }%
\providecommand \@url [1]{\endgroup\@href {#1}{\urlprefix }}%
\providecommand \urlprefix  [0]{URL }%
\providecommand \Eprint [0]{\href }%
\providecommand \doibase [0]{http://dx.doi.org/}%
\providecommand \selectlanguage [0]{\@gobble}%
\providecommand \bibinfo  [0]{\@secondoftwo}%
\providecommand \bibfield  [0]{\@secondoftwo}%
\providecommand \translation [1]{[#1]}%
\providecommand \BibitemOpen [0]{}%
\providecommand \bibitemStop [0]{}%
\providecommand \bibitemNoStop [0]{.\EOS\space}%
\providecommand \EOS [0]{\spacefactor3000\relax}%
\providecommand \BibitemShut  [1]{\csname bibitem#1\endcsname}%
\let\auto@bib@innerbib\@empty
\bibitem [{\citenamefont {Bernard}\ and\ \citenamefont
  {Krauth}(2011)}]{bernard2011two}%
  \BibitemOpen
  \bibfield  {author} {\bibinfo {author} {\bibfnamefont {E.~P.}\ \bibnamefont
  {Bernard}}\ and\ \bibinfo {author} {\bibfnamefont {W.}~\bibnamefont
  {Krauth}},\ }\href@noop {} {\bibfield  {journal} {\bibinfo  {journal} {Phys.
  Rev. Lett.}\ }\textbf {\bibinfo {volume} {107}},\ \bibinfo {pages} {155704}
  (\bibinfo {year} {2011})}\BibitemShut {NoStop}%
\bibitem [{\citenamefont {Thorneywork}\ \emph {et~al.}(2017)\citenamefont
  {Thorneywork}, \citenamefont {Abbott}, \citenamefont {Aarts},\ and\
  \citenamefont {Dullens}}]{thorneywork2017two}%
  \BibitemOpen
  \bibfield  {author} {\bibinfo {author} {\bibfnamefont {A.~L.}\ \bibnamefont
  {Thorneywork}}, \bibinfo {author} {\bibfnamefont {J.~L.}\ \bibnamefont
  {Abbott}}, \bibinfo {author} {\bibfnamefont {D.~G.}\ \bibnamefont {Aarts}}, \
  and\ \bibinfo {author} {\bibfnamefont {R.~P.}\ \bibnamefont {Dullens}},\
  }\href@noop {} {\bibfield  {journal} {\bibinfo  {journal} {Phys. Rev. Lett.}\
  }\textbf {\bibinfo {volume} {118}},\ \bibinfo {pages} {158001} (\bibinfo
  {year} {2017})}\BibitemShut {NoStop}%
\bibitem [{\citenamefont {Rosenfeld}(1989)}]{rosenfeld1989free}%
  \BibitemOpen
  \bibfield  {author} {\bibinfo {author} {\bibfnamefont {Y.}~\bibnamefont
  {Rosenfeld}},\ }\href@noop {} {\bibfield  {journal} {\bibinfo  {journal}
  {Phys. Rev. Lett.}\ }\textbf {\bibinfo {volume} {63}},\ \bibinfo {pages}
  {980} (\bibinfo {year} {1989})}\BibitemShut {NoStop}%
\bibitem [{\citenamefont {Roth}\ \emph {et~al.}(2012)\citenamefont {Roth},
  \citenamefont {Mecke},\ and\ \citenamefont {Oettel}}]{roth_2d}%
  \BibitemOpen
  \bibfield  {author} {\bibinfo {author} {\bibfnamefont {R.}~\bibnamefont
  {Roth}}, \bibinfo {author} {\bibfnamefont {K.}~\bibnamefont {Mecke}}, \ and\
  \bibinfo {author} {\bibfnamefont {M.}~\bibnamefont {Oettel}},\ }\href@noop {}
  {\bibfield  {journal} {\bibinfo  {journal} {J. Chem. Phys.}\ }\textbf
  {\bibinfo {volume} {136}},\ \bibinfo {pages} {081101} (\bibinfo {year}
  {2012})}\BibitemShut {NoStop}%
\bibitem [{\citenamefont {Thorneywork}\ \emph {et~al.}(2014)\citenamefont
  {Thorneywork}, \citenamefont {Roth}, \citenamefont {Aarts},\ and\
  \citenamefont {Dullens}}]{Thorneywork2014}%
  \BibitemOpen
  \bibfield  {author} {\bibinfo {author} {\bibfnamefont {A.~L.}\ \bibnamefont
  {Thorneywork}}, \bibinfo {author} {\bibfnamefont {R.}~\bibnamefont {Roth}},
  \bibinfo {author} {\bibfnamefont {D.~G. A.~L.}\ \bibnamefont {Aarts}}, \ and\
  \bibinfo {author} {\bibfnamefont {R.~P.~A.}\ \bibnamefont {Dullens}},\
  }\href@noop {} {\bibfield  {journal} {\bibinfo  {journal} {J. Chem. Phys.}\
  }\textbf {\bibinfo {volume} {140}},\ \bibinfo {pages} {161106} (\bibinfo
  {year} {2014})}\BibitemShut {NoStop}%
\bibitem [{\citenamefont {Zeng}\ \emph {et~al.}(1990)\citenamefont {Zeng},
  \citenamefont {Oxtoby},\ and\ \citenamefont {Rosenfeld}}]{Oxtoby1990}%
  \BibitemOpen
  \bibfield  {author} {\bibinfo {author} {\bibfnamefont {X.~C.}\ \bibnamefont
  {Zeng}}, \bibinfo {author} {\bibfnamefont {D.~W.}\ \bibnamefont {Oxtoby}}, \
  and\ \bibinfo {author} {\bibfnamefont {Y.}~\bibnamefont {Rosenfeld}},\
  }\href@noop {} {\bibfield  {journal} {\bibinfo  {journal} {Phys. Rev. A}\
  }\textbf {\bibinfo {volume} {43}},\ \bibinfo {pages} {2064} (\bibinfo {year}
  {1990})}\BibitemShut {NoStop}%
\bibitem [{\citenamefont {Xu}\ and\ \citenamefont {Baus}(1990)}]{Baus1990}%
  \BibitemOpen
  \bibfield  {author} {\bibinfo {author} {\bibfnamefont {H.}~\bibnamefont
  {Xu}}\ and\ \bibinfo {author} {\bibfnamefont {M.}~\bibnamefont {Baus}},\
  }\href@noop {} {\bibfield  {journal} {\bibinfo  {journal} {J. Phys.: Condens.
  Matter}\ }\textbf {\bibinfo {volume} {2}},\ \bibinfo {pages} {5885} (\bibinfo
  {year} {1990})}\BibitemShut {NoStop}%
\bibitem [{\citenamefont {Kranendonk}\ and\ \citenamefont
  {Frenkel}(1991)}]{Frenkel1991}%
  \BibitemOpen
  \bibfield  {author} {\bibinfo {author} {\bibfnamefont {W.~G.~T.}\
  \bibnamefont {Kranendonk}}\ and\ \bibinfo {author} {\bibfnamefont
  {D.}~\bibnamefont {Frenkel}},\ }\href@noop {} {\bibfield  {journal} {\bibinfo
   {journal} {Mol. Phys.}\ }\textbf {\bibinfo {volume} {72}},\ \bibinfo {pages}
  {679} (\bibinfo {year} {1991})}\BibitemShut {NoStop}%
\bibitem [{\citenamefont {Likos}\ and\ \citenamefont
  {Henley}(1993)}]{Likos1993}%
  \BibitemOpen
  \bibfield  {author} {\bibinfo {author} {\bibfnamefont {C.~N.}\ \bibnamefont
  {Likos}}\ and\ \bibinfo {author} {\bibfnamefont {C.~L.}\ \bibnamefont
  {Henley}},\ }\href@noop {} {\bibfield  {journal} {\bibinfo  {journal}
  {Philos. Mag. B}\ }\textbf {\bibinfo {volume} {68}},\ \bibinfo {pages} {85}
  (\bibinfo {year} {1993})}\BibitemShut {NoStop}%
\bibitem [{\citenamefont {Elder}\ \emph {et~al.}(2007)\citenamefont {Elder},
  \citenamefont {Provatas}, \citenamefont {Berry}, \citenamefont {Stefanovic},\
  and\ \citenamefont {Grant}}]{elder2007phase}%
  \BibitemOpen
  \bibfield  {author} {\bibinfo {author} {\bibfnamefont {K.}~\bibnamefont
  {Elder}}, \bibinfo {author} {\bibfnamefont {N.}~\bibnamefont {Provatas}},
  \bibinfo {author} {\bibfnamefont {J.}~\bibnamefont {Berry}}, \bibinfo
  {author} {\bibfnamefont {P.}~\bibnamefont {Stefanovic}}, \ and\ \bibinfo
  {author} {\bibfnamefont {M.}~\bibnamefont {Grant}},\ }\href@noop {}
  {\bibfield  {journal} {\bibinfo  {journal} {Phys. Rev. B}\ }\textbf {\bibinfo
  {volume} {75}},\ \bibinfo {pages} {064107} (\bibinfo {year}
  {2007})}\BibitemShut {NoStop}%
\bibitem [{\citenamefont {Greenwood}\ \emph {et~al.}(2011)\citenamefont
  {Greenwood}, \citenamefont {Ofori-Opoku}, \citenamefont {Rottler},\ and\
  \citenamefont {Provatas}}]{greenwood2011modeling}%
  \BibitemOpen
  \bibfield  {author} {\bibinfo {author} {\bibfnamefont {M.}~\bibnamefont
  {Greenwood}}, \bibinfo {author} {\bibfnamefont {N.}~\bibnamefont
  {Ofori-Opoku}}, \bibinfo {author} {\bibfnamefont {J.}~\bibnamefont
  {Rottler}}, \ and\ \bibinfo {author} {\bibfnamefont {N.}~\bibnamefont
  {Provatas}},\ }\href@noop {} {\bibfield  {journal} {\bibinfo  {journal}
  {Phys. Rev. B}\ }\textbf {\bibinfo {volume} {84}},\ \bibinfo {pages} {064104}
  (\bibinfo {year} {2011})}\BibitemShut {NoStop}%
\bibitem [{\citenamefont {Provatas}\ and\ \citenamefont
  {Majaniemi}(2010)}]{provatas2010phase}%
  \BibitemOpen
  \bibfield  {author} {\bibinfo {author} {\bibfnamefont {N.}~\bibnamefont
  {Provatas}}\ and\ \bibinfo {author} {\bibfnamefont {S.}~\bibnamefont
  {Majaniemi}},\ }\href@noop {} {\bibfield  {journal} {\bibinfo  {journal}
  {Phys. Rev. E}\ }\textbf {\bibinfo {volume} {82}},\ \bibinfo {pages} {041601}
  (\bibinfo {year} {2010})}\BibitemShut {NoStop}%
\bibitem [{\citenamefont {Oettel}\ \emph {et~al.}(2012)\citenamefont {Oettel},
  \citenamefont {Dorosz}, \citenamefont {Berghoff}, \citenamefont {Nestler},\
  and\ \citenamefont {Schilling}}]{oettel2012description}%
  \BibitemOpen
  \bibfield  {author} {\bibinfo {author} {\bibfnamefont {M.}~\bibnamefont
  {Oettel}}, \bibinfo {author} {\bibfnamefont {S.}~\bibnamefont {Dorosz}},
  \bibinfo {author} {\bibfnamefont {M.}~\bibnamefont {Berghoff}}, \bibinfo
  {author} {\bibfnamefont {B.}~\bibnamefont {Nestler}}, \ and\ \bibinfo
  {author} {\bibfnamefont {T.}~\bibnamefont {Schilling}},\ }\href@noop {}
  {\bibfield  {journal} {\bibinfo  {journal} {Phys. Rev. E}\ }\textbf {\bibinfo
  {volume} {86}},\ \bibinfo {pages} {021404} (\bibinfo {year}
  {2012})}\BibitemShut {NoStop}%
\bibitem [{\citenamefont {Russo}\ and\ \citenamefont
  {Wilding}(2017)}]{russo2017disappearance}%
  \BibitemOpen
  \bibfield  {author} {\bibinfo {author} {\bibfnamefont {J.}~\bibnamefont
  {Russo}}\ and\ \bibinfo {author} {\bibfnamefont {N.~B.}\ \bibnamefont
  {Wilding}},\ }\href@noop {} {\bibfield  {journal} {\bibinfo  {journal} {Phys.
  Rev. Lett.}\ }\textbf {\bibinfo {volume} {119}} (\bibinfo {year}
  {2017})}\BibitemShut {NoStop}%
\bibitem [{\citenamefont {Castaneda-Priego}\ \emph {et~al.}(2003)\citenamefont
  {Castaneda-Priego}, \citenamefont {Rodr{\'\i}guez-L{\'o}pez},\ and\
  \citenamefont {M{\'e}ndez-Alcaraz}}]{AO2d}%
  \BibitemOpen
  \bibfield  {author} {\bibinfo {author} {\bibfnamefont {R.}~\bibnamefont
  {Castaneda-Priego}}, \bibinfo {author} {\bibfnamefont {A.}~\bibnamefont
  {Rodr{\'\i}guez-L{\'o}pez}}, \ and\ \bibinfo {author} {\bibfnamefont
  {J.}~\bibnamefont {M{\'e}ndez-Alcaraz}},\ }\href@noop {} {\bibfield
  {journal} {\bibinfo  {journal} {J. Phys.:Condens. Matter}\ }\textbf {\bibinfo
  {volume} {15}},\ \bibinfo {pages} {S3393} (\bibinfo {year}
  {2003})}\BibitemShut {NoStop}%
\bibitem [{\citenamefont {Asakura}\ and\ \citenamefont
  {Oosawa}(1954)}]{asakura1954interaction}%
  \BibitemOpen
  \bibfield  {author} {\bibinfo {author} {\bibfnamefont {S.}~\bibnamefont
  {Asakura}}\ and\ \bibinfo {author} {\bibfnamefont {F.}~\bibnamefont
  {Oosawa}},\ }\href@noop {} {\bibfield  {journal} {\bibinfo  {journal} {J.
  Chem. Phys.}\ }\textbf {\bibinfo {volume} {22}},\ \bibinfo {pages} {1255}
  (\bibinfo {year} {1954})}\BibitemShut {NoStop}%
\bibitem [{\citenamefont {Asakura}\ and\ \citenamefont
  {Oosawa}(1958)}]{asakura1958interaction}%
  \BibitemOpen
  \bibfield  {author} {\bibinfo {author} {\bibfnamefont {S.}~\bibnamefont
  {Asakura}}\ and\ \bibinfo {author} {\bibfnamefont {F.}~\bibnamefont
  {Oosawa}},\ }\href@noop {} {\bibfield  {journal} {\bibinfo  {journal} {J.
  Polym. Sci.}\ }\textbf {\bibinfo {volume} {33}},\ \bibinfo {pages} {183}
  (\bibinfo {year} {1958})}\BibitemShut {NoStop}%
\bibitem [{\citenamefont {Schmidt}\ \emph {et~al.}(2000)\citenamefont
  {Schmidt}, \citenamefont {L{\"o}wen}, \citenamefont {Brader},\ and\
  \citenamefont {Evans}}]{schmidt2000density}%
  \BibitemOpen
  \bibfield  {author} {\bibinfo {author} {\bibfnamefont {M.}~\bibnamefont
  {Schmidt}}, \bibinfo {author} {\bibfnamefont {H.}~\bibnamefont {L{\"o}wen}},
  \bibinfo {author} {\bibfnamefont {J.~M.}\ \bibnamefont {Brader}}, \ and\
  \bibinfo {author} {\bibfnamefont {R.}~\bibnamefont {Evans}},\ }\href@noop {}
  {\bibfield  {journal} {\bibinfo  {journal} {Phys. Rev. Lett.}\ }\textbf
  {\bibinfo {volume} {85}},\ \bibinfo {pages} {1934} (\bibinfo {year}
  {2000})}\BibitemShut {NoStop}%
\bibitem [{\citenamefont {Mortazavifar}\ and\ \citenamefont
  {Oettel}(2016)}]{mortazavifar2016fundamental}%
  \BibitemOpen
  \bibfield  {author} {\bibinfo {author} {\bibfnamefont {M.}~\bibnamefont
  {Mortazavifar}}\ and\ \bibinfo {author} {\bibfnamefont {M.}~\bibnamefont
  {Oettel}},\ }\href@noop {} {\bibfield  {journal} {\bibinfo  {journal} {J.
  Phys.:Condens. Matter}\ }\textbf {\bibinfo {volume} {28}},\ \bibinfo {pages}
  {244018} (\bibinfo {year} {2016})}\BibitemShut {NoStop}%
\bibitem [{\citenamefont {Tarazona}\ and\ \citenamefont
  {Rosenfeld}(1997)}]{Tarazona1997}%
  \BibitemOpen
  \bibfield  {author} {\bibinfo {author} {\bibfnamefont {P.}~\bibnamefont
  {Tarazona}}\ and\ \bibinfo {author} {\bibfnamefont {Y.}~\bibnamefont
  {Rosenfeld}},\ }\href@noop {} {\bibfield  {journal} {\bibinfo  {journal}
  {Phys. Rev. E}\ }\textbf {\bibinfo {volume} {55}},\ \bibinfo {pages} {4873}
  (\bibinfo {year} {1997})}\BibitemShut {NoStop}%
\bibitem [{\citenamefont {Tarazona}(2000)}]{Tarazona2000}%
  \BibitemOpen
  \bibfield  {author} {\bibinfo {author} {\bibfnamefont {P.}~\bibnamefont
  {Tarazona}},\ }\href@noop {} {\bibfield  {journal} {\bibinfo  {journal}
  {Phys. Rev. Lett.}\ }\textbf {\bibinfo {volume} {84}},\ \bibinfo {pages}
  {694} (\bibinfo {year} {2000})}\BibitemShut {NoStop}%
\bibitem [{\citenamefont {Oettel}\ \emph {et~al.}(2010)\citenamefont {Oettel},
  \citenamefont {G{\"o}rig}, \citenamefont {H{\"a}rtel}, \citenamefont
  {L{\"o}wen}, \citenamefont {Radu},\ and\ \citenamefont
  {Schilling}}]{oettel2010free}%
  \BibitemOpen
  \bibfield  {author} {\bibinfo {author} {\bibfnamefont {M.}~\bibnamefont
  {Oettel}}, \bibinfo {author} {\bibfnamefont {S.}~\bibnamefont {G{\"o}rig}},
  \bibinfo {author} {\bibfnamefont {A.}~\bibnamefont {H{\"a}rtel}}, \bibinfo
  {author} {\bibfnamefont {H.}~\bibnamefont {L{\"o}wen}}, \bibinfo {author}
  {\bibfnamefont {M.}~\bibnamefont {Radu}}, \ and\ \bibinfo {author}
  {\bibfnamefont {T.}~\bibnamefont {Schilling}},\ }\href@noop {} {\bibfield
  {journal} {\bibinfo  {journal} {Phys. Rev. E}\ }\textbf {\bibinfo {volume}
  {82}},\ \bibinfo {pages} {051404} (\bibinfo {year} {2010})}\BibitemShut
  {NoStop}%
\bibitem [{\citenamefont {H\"artel}\ \emph {et~al.}(2012)\citenamefont
  {H\"artel}, \citenamefont {Oettel}, \citenamefont {Rozas}, \citenamefont
  {Egelhaaf}, \citenamefont {Horbach},\ and\ \citenamefont
  {L\"owen}}]{Haertel2012}%
  \BibitemOpen
  \bibfield  {author} {\bibinfo {author} {\bibfnamefont {A.}~\bibnamefont
  {H\"artel}}, \bibinfo {author} {\bibfnamefont {M.}~\bibnamefont {Oettel}},
  \bibinfo {author} {\bibfnamefont {R.~E.}\ \bibnamefont {Rozas}}, \bibinfo
  {author} {\bibfnamefont {S.~U.}\ \bibnamefont {Egelhaaf}}, \bibinfo {author}
  {\bibfnamefont {J.}~\bibnamefont {Horbach}}, \ and\ \bibinfo {author}
  {\bibfnamefont {H.}~\bibnamefont {L\"owen}},\ }\href@noop {} {\bibfield
  {journal} {\bibinfo  {journal} {Phys. Rev. Lett.}\ }\textbf {\bibinfo
  {volume} {108}},\ \bibinfo {pages} {226101} (\bibinfo {year}
  {2012})}\BibitemShut {NoStop}%
\bibitem [{\citenamefont {Wittmann}\ \emph {et~al.}(2017)\citenamefont
  {Wittmann}, \citenamefont {Sitta}, \citenamefont {Smallenburg},\ and\
  \citenamefont {L\"owen}}]{Wittmann2017}%
  \BibitemOpen
  \bibfield  {author} {\bibinfo {author} {\bibfnamefont {R.}~\bibnamefont
  {Wittmann}}, \bibinfo {author} {\bibfnamefont {C.~E.}\ \bibnamefont {Sitta}},
  \bibinfo {author} {\bibfnamefont {F.}~\bibnamefont {Smallenburg}}, \ and\
  \bibinfo {author} {\bibfnamefont {H.}~\bibnamefont {L\"owen}},\ }\href@noop
  {} {\bibfield  {journal} {\bibinfo  {journal} {J. Chem. Phys.}\ }\textbf
  {\bibinfo {volume} {147}},\ \bibinfo {pages} {134908} (\bibinfo {year}
  {2017})}\BibitemShut {NoStop}%
\bibitem [{\citenamefont {Brader}\ \emph {et~al.}(2003)\citenamefont {Brader},
  \citenamefont {Evans},\ and\ \citenamefont {Schmidt}}]{Brader2003}%
  \BibitemOpen
  \bibfield  {author} {\bibinfo {author} {\bibfnamefont {J.~M.}\ \bibnamefont
  {Brader}}, \bibinfo {author} {\bibfnamefont {R.}~\bibnamefont {Evans}}, \
  and\ \bibinfo {author} {\bibfnamefont {M.}~\bibnamefont {Schmidt}},\
  }\href@noop {} {\bibfield  {journal} {\bibinfo  {journal} {Mol. Phys.}\
  }\textbf {\bibinfo {volume} {101}},\ \bibinfo {pages} {3349} (\bibinfo {year}
  {2003})}\BibitemShut {NoStop}%
\bibitem [{\citenamefont {Cox}\ and\ \citenamefont
  {Matthews}(2002)}]{cox2002exponential}%
  \BibitemOpen
  \bibfield  {author} {\bibinfo {author} {\bibfnamefont {S.~M.}\ \bibnamefont
  {Cox}}\ and\ \bibinfo {author} {\bibfnamefont {P.~C.}\ \bibnamefont
  {Matthews}},\ }\href@noop {} {\bibfield  {journal} {\bibinfo  {journal} {J.
  Comput. Phys.}\ }\textbf {\bibinfo {volume} {176}},\ \bibinfo {pages} {430}
  (\bibinfo {year} {2002})}\BibitemShut {NoStop}%
\bibitem [{\citenamefont {Adland}\ \emph {et~al.}(2013)\citenamefont {Adland},
  \citenamefont {Karma}, \citenamefont {Spatschek}, \citenamefont {Buta},\ and\
  \citenamefont {Asta}}]{adland2013phase}%
  \BibitemOpen
  \bibfield  {author} {\bibinfo {author} {\bibfnamefont {A.}~\bibnamefont
  {Adland}}, \bibinfo {author} {\bibfnamefont {A.}~\bibnamefont {Karma}},
  \bibinfo {author} {\bibfnamefont {R.}~\bibnamefont {Spatschek}}, \bibinfo
  {author} {\bibfnamefont {D.}~\bibnamefont {Buta}}, \ and\ \bibinfo {author}
  {\bibfnamefont {M.}~\bibnamefont {Asta}},\ }\href@noop {} {\bibfield
  {journal} {\bibinfo  {journal} {Phys. Rev. B}\ }\textbf {\bibinfo {volume}
  {87}},\ \bibinfo {pages} {024110} (\bibinfo {year} {2013})}\BibitemShut
  {NoStop}%
\bibitem [{\citenamefont {Stopper}\ and\ \citenamefont {Roth}(2017)}]{GPU_FMT}%
  \BibitemOpen
  \bibfield  {author} {\bibinfo {author} {\bibfnamefont {D.}~\bibnamefont
  {Stopper}}\ and\ \bibinfo {author} {\bibfnamefont {R.}~\bibnamefont {Roth}},\
  }\href@noop {} {\bibfield  {journal} {\bibinfo  {journal} {J. Chem. Phys.}\
  }\textbf {\bibinfo {volume} {147}},\ \bibinfo {pages} {064508} (\bibinfo
  {year} {2017})}\BibitemShut {NoStop}%
\bibitem [{\citenamefont {Nickolls}\ \emph {et~al.}(2008)\citenamefont
  {Nickolls}, \citenamefont {Buck}, \citenamefont {Garland},\ and\
  \citenamefont {Skadron}}]{CUDA}%
  \BibitemOpen
  \bibfield  {author} {\bibinfo {author} {\bibfnamefont {J.}~\bibnamefont
  {Nickolls}}, \bibinfo {author} {\bibfnamefont {I.}~\bibnamefont {Buck}},
  \bibinfo {author} {\bibfnamefont {M.}~\bibnamefont {Garland}}, \ and\
  \bibinfo {author} {\bibfnamefont {K.}~\bibnamefont {Skadron}},\ }\href@noop
  {} {\bibfield  {journal} {\bibinfo  {journal} {ACM Queue}\ }\textbf {\bibinfo
  {volume} {6}},\ \bibinfo {pages} {40} (\bibinfo {year} {2008})}\BibitemShut
  {NoStop}%
\bibitem [{\citenamefont {Oettel}(2012)}]{oettel2012mode}%
  \BibitemOpen
  \bibfield  {author} {\bibinfo {author} {\bibfnamefont {M.}~\bibnamefont
  {Oettel}},\ }\href@noop {} {\bibfield  {journal} {\bibinfo  {journal} {J.
  Phys.:Condens. Matter}\ }\textbf {\bibinfo {volume} {24}},\ \bibinfo {pages}
  {464124} (\bibinfo {year} {2012})}\BibitemShut {NoStop}%
\bibitem [{\citenamefont {Gonzalez}(2016)}]{Gonzalez2016}%
  \BibitemOpen
  \bibfield  {author} {\bibinfo {author} {\bibfnamefont {A.~E.}\ \bibnamefont
  {Gonzalez}},\ }\href@noop {} {\bibfield  {journal} {\bibinfo  {journal}
  {Crystals}\ }\textbf {\bibinfo {volume} {6}},\ \bibinfo {pages} {46}
  (\bibinfo {year} {2016})}\BibitemShut {NoStop}%
\bibitem [{\citenamefont {Amini}\ and\ \citenamefont
  {Laird}(2008)}]{Laird2008}%
  \BibitemOpen
  \bibfield  {author} {\bibinfo {author} {\bibfnamefont {M.}~\bibnamefont
  {Amini}}\ and\ \bibinfo {author} {\bibfnamefont {B.~B.}\ \bibnamefont
  {Laird}},\ }\href@noop {} {\bibfield  {journal} {\bibinfo  {journal} {Phys.
  Rev. B}\ }\textbf {\bibinfo {volume} {78}},\ \bibinfo {pages} {144112}
  (\bibinfo {year} {2008})}\BibitemShut {NoStop}%
\bibitem [{\citenamefont {Becker}\ \emph {et~al.}(2009)\citenamefont {Becker},
  \citenamefont {Olmsted}, \citenamefont {Asta}, \citenamefont {Hoyt},\ and\
  \citenamefont {Foiles}}]{Foiles2009}%
  \BibitemOpen
  \bibfield  {author} {\bibinfo {author} {\bibfnamefont {C.~A.}\ \bibnamefont
  {Becker}}, \bibinfo {author} {\bibfnamefont {D.~L.}\ \bibnamefont {Olmsted}},
  \bibinfo {author} {\bibfnamefont {M.}~\bibnamefont {Asta}}, \bibinfo {author}
  {\bibfnamefont {J.~J.}\ \bibnamefont {Hoyt}}, \ and\ \bibinfo {author}
  {\bibfnamefont {S.~M.}\ \bibnamefont {Foiles}},\ }\href@noop {} {\bibfield
  {journal} {\bibinfo  {journal} {Phys. Rev. B}\ }\textbf {\bibinfo {volume}
  {79}},\ \bibinfo {pages} {054109} (\bibinfo {year} {2009})}\BibitemShut
  {NoStop}%
\bibitem [{\citenamefont {Vink}\ \emph {et~al.}(2011)\citenamefont {Vink},
  \citenamefont {Neuhaus},\ and\ \citenamefont {L{\"o}wen}}]{vink2011fluid}%
  \BibitemOpen
  \bibfield  {author} {\bibinfo {author} {\bibfnamefont {R.~L.}\ \bibnamefont
  {Vink}}, \bibinfo {author} {\bibfnamefont {T.}~\bibnamefont {Neuhaus}}, \
  and\ \bibinfo {author} {\bibfnamefont {H.}~\bibnamefont {L{\"o}wen}},\
  }\href@noop {} {\bibfield  {journal} {\bibinfo  {journal} {J. Chem. Phys.}\
  }\textbf {\bibinfo {volume} {134}},\ \bibinfo {pages} {204907} (\bibinfo
  {year} {2011})}\BibitemShut {NoStop}%
\bibitem [{\citenamefont {Vink}\ and\ \citenamefont
  {Horbach}(2004)}]{vink2004grand}%
  \BibitemOpen
  \bibfield  {author} {\bibinfo {author} {\bibfnamefont {R.}~\bibnamefont
  {Vink}}\ and\ \bibinfo {author} {\bibfnamefont {J.}~\bibnamefont {Horbach}},\
  }\href@noop {} {\bibfield  {journal} {\bibinfo  {journal} {J. Chem. Phys.}\
  }\textbf {\bibinfo {volume} {121}},\ \bibinfo {pages} {3253} (\bibinfo {year}
  {2004})}\BibitemShut {NoStop}%
\end{thebibliography}

%


\end{document}